\documentclass[aps,prl,twocolumn,showpacs,superscriptaddress,groupedaddress,preprintnumbers,bm,amsmath,amssymb,amsfonts,showpacs,superscriptaddress]{revtex4-1}

\usepackage[hyperindex,breaklinks]{hyperref}
\usepackage{graphicx,epstopdf}

\usepackage{natbib,bm}
\usepackage[usenames,dvipsnames]{color}
\usepackage{slashed}    
\usepackage{hyperref}
\usepackage{breakurl}
\usepackage{multirow}
\usepackage{bbold}

\def\be{\begin{equation}}
\def\ee{\end{equation}}
\def\ba{\begin{eqnarray}}
\def\ea{\end{eqnarray}}
\def\ge{\mathrel{\raise.3ex\hbox{$>$\kern-.75em\lower1ex\hbox{$\sim$}}}}
\def\la{\mathrel{\raise.3ex\hbox{$<$\kern-.75em\lower1ex\hbox{$\sim$}}}}

\def\simgt{\mathrel{\raise.3ex\hbox{$>$\kern-.75em\lower1ex\hbox{$\sim$}}}}
\def\simlt{\mathrel{\raise.3ex\hbox{$<$\kern-.75em\lower1ex\hbox{$\sim$}}}}

\newcommand{\nc}{\newcommand}

\nc{\gone}{\bar g_{\pi NN}^{(1)}}
\nc{\gzero}{\bar g_{\pi NN}^{(0)}}
\nc{\al}{\alpha}
\nc{\ga}{\gamma}
\nc{\de}{\delta}
\nc{\ep}{\epsilon}
\nc{\ze}{\zeta}
\nc{\et}{\eta}
\nc{\ka}{\kappa}
\nc{\rh}{\rho}
\nc{\si}{\sigma}
\nc{\ta}{\tau}
\nc{\up}{\upsilon}
\nc{\ph}{\phi}
\nc{\ch}{\chi}
\nc{\ps}{\psi}
\nc{\om}{\omega}
\nc{\Ga}{\Gamma}
\nc{\De}{\Delta}
\nc{\La}{\Lambda}
\nc{\Up}{\Upsilon}
\nc{\Ph}{\Phi}
\nc{\Ps}{\Psi}
\nc{\Om}{\Omega}
\nc{\ptl}{\partial}
\nc{\del}{\nabla}
\nc{\ov}{\overline}
\nc{\newcaption}[1]{\centerline{\parbox{15cm}{\caption{#1}}}}
\nc{\us}{U(1)$_S$}

\def\beq{\begin{equation}}
\def\eeq{\end{equation}}
\def\bmat{\begin{displaymath}}
\def\emat{\end{displaymath}}
\def\bear{\begin{eqnarray}}
\def\eear{\end{eqnarray}}
\def\ba{\begin{eqnarray}}
\def\ea{\end{eqnarray}}
\def\bery{\begin{array}}
\def\ery{\end{array}}
\def\bit{\begin{itemize}}
\def\eit{\end{itemize}}
\def\ben{\begin{enumerate}}
\def\een{\end{enumerate}}
\def\btab{\begin{tabular}}
\def\etab{\end{tabular}}
\def\btbl{\begin{table}}
\def\etbl{\end{table}}
\def\bfig{\begin{figure}[htb]}
\def\efig{\end{figure}}
\def\bpic{\begin{picture}}
\def\epic{\end{picture}}

\def\ga{\mathrel{\raise.3ex\hbox{$>$\kern-.75em\lower1ex\hbox{$\sim$}}}}
\def\la{\mathrel{\raise.3ex\hbox{$<$\kern-.75em\lower1ex\hbox{$\sim$}}}}
\def\gappeq{\mathrel{\rlap {\raise.5ex\hbox{$>$}}
{\lower.5ex\hbox{$\sim$}}}}
\def\lappeq{\mathrel{\rlap{\raise.5ex\hbox{$<$}}
{\lower.5ex\hbox{$\sim$}}}}

\def\gyr{{\rm \, G\kern-0.125em yr}}
\def\mev{{\rm \, Me\kern-0.125em V}}
\def\gev{{\rm \, Ge\kern-0.125em V}}
\def\tev{{\rm \, Te\kern-0.125em V}}

\newcommand{\Sone}{\ensuremath{\mathrm{S}1}}
\newcommand{\Stwo}{\ensuremath{\mathrm{S}2}}
\newcommand{\Si}{\ensuremath{\mathrm{S}i}}
\newcommand{\pdf}{\ensuremath{\mathrm{pdf}}}
\newcommand{\PE}{\ensuremath{\mathrm{PE}}}

\newcommand{\eV}{\ensuremath{\mathrm{eV}}}
\newcommand{\keV}{\ensuremath{\mathrm{keV}}}

\DeclareMathOperator{\binomial}{binom}
\DeclareMathOperator{\gaussian}{gauss}

\begin{document}
 
\preprint{CALT-2017-042}

\title{Directly Detecting MeV-scale Dark Matter via Solar Reflection}

\author{Haipeng An}
\affiliation{Walter Burke Institute for Theoretical Physics, California Institute of Technology, Pasadena, CA, 91125, USA}
\affiliation{Department of Physics, Tsinghua University, Beijing 100084, China}

\author{Maxim Pospelov}
\affiliation{Department of Physics and Astronomy, University of Victoria, 
Victoria, BC V8P 5C2, Canada}
\affiliation{Perimeter Institute for Theoretical Physics, Waterloo, ON N2J 2W9, 
Canada}

\author{Josef Pradler}
\affiliation{Institute of High Energy Physics, Austrian Academy of Sciences, 1050 Vienna, Austria}

\author{Adam Ritz}
\affiliation{Department of Physics and Astronomy, University of Victoria, 
Victoria, BC V8P 5C2, Canada}

\date{August 2017}

\begin{abstract}
\noindent 
If dark matter (DM) particles are lighter than a few MeV$/c^2$ and can scatter off electrons, their interaction within the solar interior results in a considerable hardening of the spectrum of galactic dark matter received on Earth. For a large range of the mass vs~cross section parameter space, $\{m_e, \sigma_e\}$, 
the  `reflected' component of the DM flux is far more energetic than the endpoint of the ambient galactic DM energy distribution, making it detectable with 
existing DM detectors sensitive to an energy deposition of $10-10^3$~eV. After numerically simulating the small reflected component of the 
DM flux, we calculate its subsequent signal due to scattering on detector electrons, deriving new constraints on $\sigma_e$ in the MeV and sub-MeV range using existing data from the XENON10/100, LUX, PandaX-II, and XENON1T experiments, as well as making projections for future low threshold direct detection experiments. 
\end{abstract}
\maketitle

\paragraph{Introduction.} Astrophysics and cosmology provide one of the strongest arguments for an extension to the Standard Model (SM) of particle physics, through the need for dark matter (DM). 
The `theory-space' for dark matter remains vast, motivating a range of
experimental approaches. A well-motivated class of models achieve the
required relic abundance through thermal freeze-out during the early
radiation-dominated epoch, which points to particles with weak-scale
interactions -- weakly interacting massive particles (WIMPs) -- with
the required annihilation rate
$\langle \si_{\rm ann} v\rangle \sim 10^{-36}\,{\rm cm^2}$ ($c=1$ from now on). A range of direct
detection experiments, searching for the elastic scattering of such DM
particles in the galactic halo on nuclei, have now pushed the limit down to
the scale of $\sigma_n \sim 10^{-46}\,$cm$^2$ for weak-scale masses
\cite{Aprile:2017iyp}.

Since cold DM in the halo is non-relativisitic, detector thresholds
ensure that the sensitivity weakens dramatically for masses below a
few
GeV~\cite{Aprile:2017iyp,Akerib:2016vxi,Tan:2016zwf,Angloher:2015ewa,Kouvaris:2016afs,McCabe:2017rln}. In
recent years, this has motivated efforts to extend this reach to lower
mass scales that still allow for viable thermal relic DM candidates
(see \textit{e.g.}~\cite{Battaglieri:2017aum,Alexander:2016aln}),
often with interactions mediated by new light (dark) forces
\cite{Boehm:2003hm}. These efforts have included searches at
colliders, fixed target proton and electron experiments, and also
consideration of direct detection via electron scattering
\cite{Batell:2009di,deNiverville:2011it,deNiverville:2012ij,Bjorken:2009mm,
  Kahn:2014sra,Dobrescu:2014ita,Izaguirre:2013uxa,Izaguirre:2014dua,Batell:2014mga,Essig:2011nj,Essig:2012yx}. The
latter approach offers the possibility of extending conventional
direct detection down to masses of $\sim 10\,$MeV
\cite{Essig:2011nj,Essig:2012yx,Essig:2015cda}, where the halo DM
kinetic energy is
$E_{\rm DM}^{\rm halo}\sim \frac{1}{2}m_{\rm DM} v^2\sim
5\,$eV. Lowering the energy threshold by $O(10)$ down to $1~{\rm eV}$
appears feasible \cite{Battaglieri:2017aum}, and there are theoretical
proposals for more significant reductions (see {\em e.g.}
\cite{Hochberg:2015pha}).

\begin{figure}
  \includegraphics[viewport=72 555 540 730, clip=true,
  scale=0.4,width=0.45\textwidth]{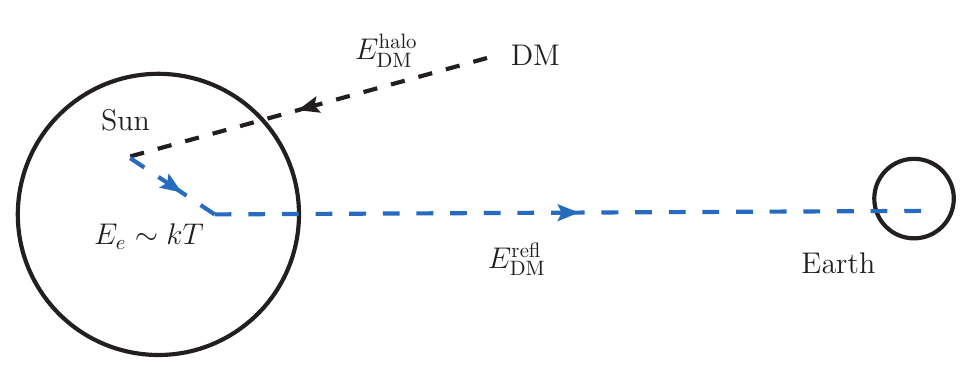}
\caption{\footnotesize A schematic illustration of the reflected dark
  matter flux generated through solar scattering. For bound solar
  electrons with energy $E_e\sim kT$, the DM recoil energy is bounded
  by the expression in Eq.~(\ref{Erec}) and can be $\sim\,$keV.
  }
\label{fig1}
\end{figure}

\begin{figure}
 \includegraphics[width=\columnwidth]{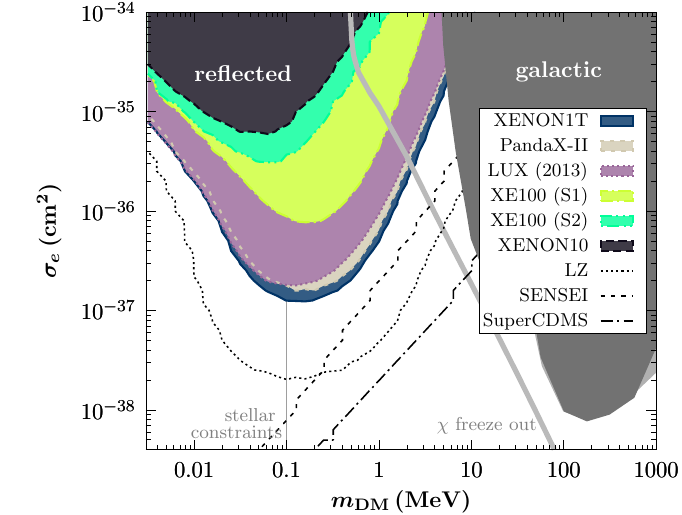}
 \caption{\footnotesize Exclusion contours for 'reflected DM' from a
   range of experiments are shown in comparison to previous limits
   from XENON10 and XENON100 on scattering from the 'galactic DM' halo
   population~\cite{Essig:2017kqs,Essig:2012yx}. Filled contours
   reflect current limits, while dashed contours denote future
   projections. The thick gray relic density contour is for the DM
   model in Eq.~(\ref{eq:model}).  A vertical line at 100~keV
   indicates a schematic lower limit on $m_{\rm DM}$ from stellar
   energy loss while the more model-dependent cosmological
   $N_{\rm eff}$ constraint is not shown (see text).}
 \label{results}
\end{figure}

In this Letter, we point out that further direct detection sensitivity
to DM in the 10~keV -- 10~MeV mass range is possible through
consideration of `reflected DM' initially scattered by more energetic
electrons in the Sun (or the Earth) prior to scattering in the
detector. This double (or multiple) scattering trajectory allows the
DM kinetic energy to be lifted to the keV range. Depending on the value of the 
reduced DM-$e$ mass, $\mu_{{\rm DM},e}$, a single scatter 
may result in the energy of the reflected DM, 
\be E_{\rm
  DM}^{\rm refl} < E_{\rm DM}^{\rm refl,max} = 
\frac{4 E_e\mu_{{\rm DM},e}}{m_e+m_{\rm DM}}= \frac{4 E_e m_{{\rm DM}} m_e}{(m_e+m_{\rm DM})^2}, \label{Erec} \ee 
being much higher than $E_{\rm DM}^{\rm halo}$ and indeed comparable to the typical solar electron kinetic energy $E_e\sim kT_e \sim O({\rm keV})$.
Thus $E_{\rm DM}^{\rm refl}$ can be
above the detection threshold for a number of existing experiments,
including XENON10, XENON100, LUX, PandaX-II and XENON1T.

  The basic scenario is summarized in
Fig.~1. DM scattering off free electrons in the Sun
 generates a
new (more energetic) component of the flux impinging on the
Earth. While there is necessarily a geometric suppression factor,
associated with re-scattering in the direction of the Earth, we find
that this is still sufficient to produce new levels of sensitivity to
MeV and sub-MeV dark matter, where 
no direct detection constraints previously existed. 
The limits and projected sensitivity from
electron scattering at a number of experiments are summarized in
Fig.~\ref{results}.

\paragraph{ Solar Reflection of Light DM.} 

DM scattering on particles inside the Sun has been extensively studied
as an ingredient for the indirect signature of DM annihilation to high
energy neutrinos.  The evolution of DM that intercepts the Sun depends
crucially on its mass.  Given a large enough elastic cross section on
nuclei, WIMP dark matter with mass above a few GeV can be efficiently
captured and thermalized.  However, for light DM, the capture process
is less efficient, and DM tends to re-scatter at larger radii and
evaporate.  The `evaporated' component of the DM flux impinging on the
Earth may help improve sensitivity to $\sigma_n$
\cite{Kouvaris:2015nsa}, and, as we are going to show, the effect
mediated by $\sigma_e$ is even more pronounced for MeV and sub-MeV
mass reflected DM; for a detailed comparison between DM scattering
on electrons vs.~nucleons inside the sun see~\cite{Garani:2017jcj}.

Depending on the scattering cross
section $\si_e$, and thus the mean free path, reflection may occur after
just one or two interactions, or after partial thermalization through
multiple scatters within the Sun. The reflected DM flux will be
determined via a simulation which tracks the kinematics after initial
entry into the Sun. 
We will assume a velocity-independent $s$-wave cross section, but it is notable that
the relative importance of the reflected flux would be enhanced for
models with a power-like dependence of the cross section on the relative electron-DM velocity, $\si_e \propto (v_{\rm rel})^n$,
such as would occur {\em e.g.} for scattering via higher multipoles.
We note in passing that energy-loss or transfer from or inside the sun due to the scattering is negligible for the considered parameter region.

To determine the reflected contribution to the DM flux, the incoming
velocity is assumed to follow a Maxwell-Boltzmann distribution with an
expectation value of $10^{-3}$, and an escape velocity cut-off at
$2\times10^{-3}$. This velocity is negligible compared to solar
electrons, and thus DM that scatters in the Sun acquires
$E_{\rm DM}^{\rm recoil}\sim T$. To gain some intuition, we note first
that the probability of scattering off electrons in the solar core is
approximately
$\si_e \times R_{\rm core} \times n_e^{\rm core} \sim \si_e/{\rm pb}$,
and thus the Sun scatters efficiently if $\si_e \gg 10^{-36}\,{\rm cm^2}$. In
this optically thick regime, scattering occurs in the convective
zone at a characteristic radius $R_{\rm scatt}$ given implicitly by
$\si_e \int_{R_{\rm scatt}}^{R_\odot} n_e(R)dR\sim {\cal O}(1)$. It
follows that the electron temperature, and thus the recoil energy,
will depend on $\si_e$ which in turn determines $R_{\rm scatt}$, 
through the radius-temperature relation \cite{Bahcall:2004pz}. As the cross section
is reduced, $R_{\rm scatt}$ also decreases and $E^{\rm refl,max}_{\rm DM}$
increases as scattering occurs in hotter regions of the
core. Further decreasing the cross section ultimately increases 
the mean free path $\sim (\sigma_en_e)^{-1}$  beyond the solar radius, and the strength of the reflected flux is suppressed. 
The scattering probability and the background DM flux in the halo, defined through the number density 
and average velocity as  $\Ph^{\rm halo} \equiv n_{\rm DM} v_{\rm DM}^{\rm halo} $, may be 
combined into a simple estimate for the reflected DM flux incident on the Earth,
\be
 \Ph_{\rm refl} \sim \frac{\Ph^{\rm halo}}{4}  \times \left\{ \begin{array}{ll} \frac{4S_g}{3}\left(\frac{R_{\rm core}}{1\,{\rm A.U.}}\right)^2 \si_e n_e^{\rm core} R_{\rm core}, & \si_e \ll 1\,{\rm pb}, \\
                        S_g\left(\frac{R_{\rm scatt}}{1\,{\rm A.U.}}\right)^2, & \si_e \gg 1\,{\rm pb}. \end{array} \right. \label{flux}
\ee
In the estimate (\ref{flux}), the overall coefficient of $1/4$ has a geometric origin from $\pi R_\odot^2/(4\pi(1\,{\rm A.U.})^2)$. 
$S_g$ denotes the gravitational focussing effect that enhances the area at spatial infinity subtended by the effective solar scattering disk $\pi R^2_{\rm scatt}$. For example, at 
$R_{\rm scatt}\sim R_\odot$, we have $S_g \sim 1+v^2_{\rm esc}/(v_{\rm DM}^{\rm halo} )^2 \sim O(10) $,
given the value of the solar escape velocity $v_{\rm esc}$. We note that the overall energy extracted from the Sun by reflected DM does not 
exceed $\sim 10 T\times\pi  R_\odot^2\Ph^{\rm halo}$, and therefore is not constrained by solar energetics being many orders of magnitude below solar luminosity.

Taking a representative choice of $m_{\rm DM}\sim 3\,$MeV, 
one can estimate the maximum value of the recoil energy distribution to be 
$\sim 0.5 \,T(R_{\rm scatt})$ at
$\si_e \gg 10^{-36}{\rm cm^2}$.  For example, a single scatter would accelerate a 3 MeV DM particle up to $\sim 100$~eV energy for
$\si_e \sim 10^{-33}\ \rm cm^2$ ($R_{\rm scatt}{=}0.8R_\odot$). The reflected flux (\ref{flux}) in this optically thick regime is $10^{5}\,$cm$^{-2}$s$^{-1}$,
leading to ~${\cal O}(20)$ ionizations/day in 1kg of Xe. This constitutes a detectable signal, and motivates a more detailed analysis.

Our preliminary estimates (\ref{flux}) need to be augmented to 
include the possibility of multiple scattering, which can significantly impact the energy of the reflected particles. 
Since this is  difficult to treat analytically, we will make use of a simulation to determine the energy spectrum and intensity of the reflected DM flux.
The simulation scans the initial velocity and impact parameter to determine the initial trajectory into the Sun. The step size was chosen as $0.01R_\odot$, and the Standard Solar Model \cite{Bahcall:2004pz} was used to determine the temperature, density and elemental abundance at each given radius. For a given cross section $\si_e$, the scattering rate was then determined probabilistically. If DM does not scatter, it propagates to the next step with velocity shifted according to the gravitational potential. If DM scatters, the electron momentum was generated according to the temperature distribution, and the new trajectory determined by first boosting to the DM-electron rest frame, and assuming an $s$-wave cross section. The gravitational effect on the trajectory was included after each nontrivial scattering. This process was repeated until the DM particle exits the Sun. 

\begin{figure}
 \centerline{
 \includegraphics[width=0.45\textwidth]{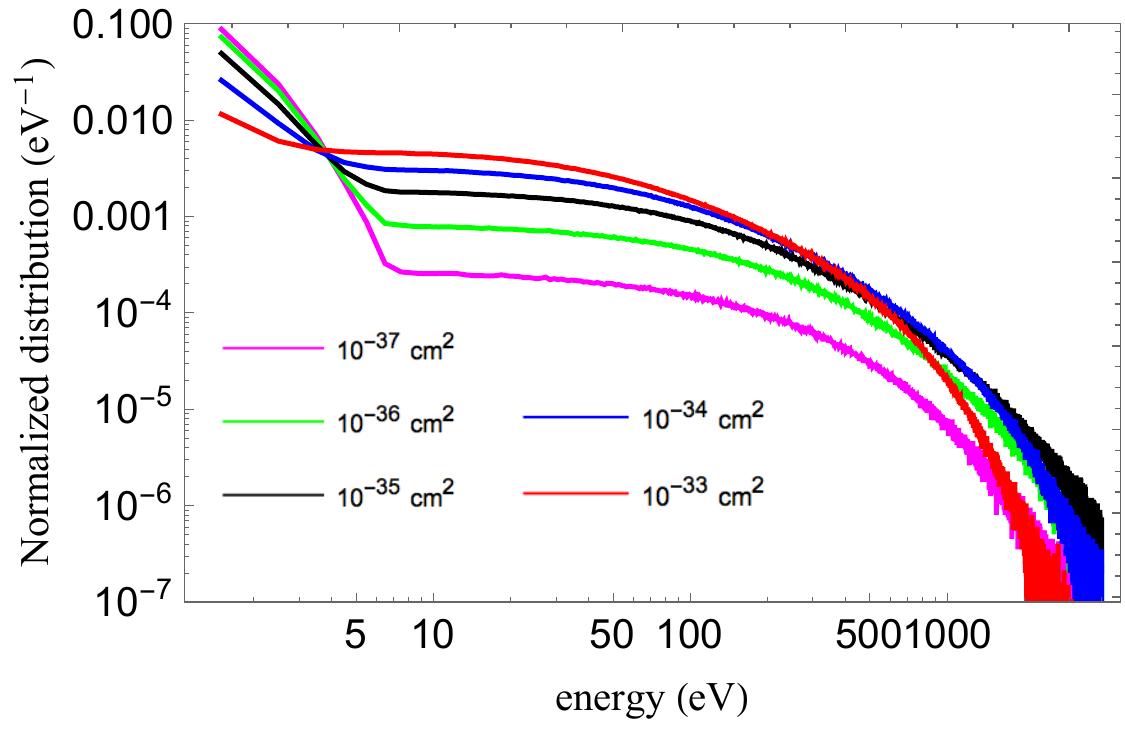}
}
 \caption{\footnotesize Normalized energy distributions $F_{A_\rho= 16\pi R^2_{\odot}}(E)$ (in $eV$),  are shown for reflected DM with a mass of 3~MeV and the range of scattering cross sections indicated. The initial velocity is assumed to follow a Maxwell-Boltzmann distribution with an expectation value of $10^{-3}$, and an escape velocity cut-off at $2\times10^{-3}$. It is apparent that the distributions below 5-7~eV tend to that of the background halo.}
 \label{fig3}
\end{figure}

We find that it is sufficient to limit our simulations by a maximal impact parameter $\rho_{\rm max}  = 4R_\odot$. Outside that range, only the slowest DM particles will enter the Sun, giving a highly subdominant contribution to the reflected flux. Thus, we simulate the energy distribution $F_{A_\rho}(E)$ of particles interacting with (or missing) 
the Sun initially collected from the $A_\rho = 16 \pi R_\odot^2$ impact area. 
After accounting for the gravitational redshift,  $E\to E - m_{\rm DM}v_{\rm esc}^2/2$, the 
distribution is normalized to unity, $\int_{0}^{\infty} dE F_{A_\rho}(E) = 1$, and the resulting reflected DM flux at Earth determined via 
\begin{equation}
\label{diffFlux}
 \frac{d \Ph_{\rm refl}}{dE} = \Ph_{\rm halo} \times \frac{A_\rho F_{A_\rho}(E)}{4 \pi (1\,{\rm A.U.})^2} .
\end{equation}
As there is some arbitrariness in $A_\rho$, the simulated reflected flux contains an admixture of the initial un-scattered distribution. 
This does not affect subsequent calculations because this component stays below detection thresholds.

Fig.~3 shows the final kinetic energy distribution at Earth for 3~MeV DM particles. 
For $\si_e \sim 1\,$nb, the
distribution turns over close to 100~eV, consistent with naive estimates. 
Moreover, tracking the trajectories indicates that DM
does indeed have a higher probability to enter the core region if the
cross section is below about $10^{-34}\,$cm$^2$. Despite the lower
cross-section, the enhanced core temperature can in turn lead to less
scatters for DM to exit the Sun, resulting in the observed enhancement
in the tail of the distribution as the cross-section
decreases. However, the effect eventually turns off once the cross
section drops well below a pb, as the mean free path and thus the
collision rate becomes too low.

\paragraph{ Direct detection via electron scattering.} With the reflected DM
flux and velocity distribution in hand, the scattering signatures can be
determined along the lines of the DM-electron scattering analysis
of~\cite{Essig:2011nj,Essig:2012yx}, with the
modifications outlined below.
We consider DM scattering off bound electrons in the detector, having
fixed energy $E_e = m_e - |E_B|$, with binding energy $E_B$ and a range
of momenta. The process of interest corresponds to atomic ionization
${\rm DM} + A \rightarrow {\rm DM} + A^+ + e^-$ with DM three-momentum transfer
$\vec q$.  To match the literature, we write the differential
scattering rate as a function of electron recoil energy in terms of a
reference cross-section~$\sigma_e$~\cite{Essig:2012yx},
\begin{align}
\label{eq:ddrate}
\frac{  d \langle \sigma_{nl} v \rangle }{ d\ln E_{R,e} } =  
 \frac{  \sigma_e }{8 \mu_{\rm DM, e}^2} 
  \int dq \, q |f_{nl}( q, p_e')|^2  |F_{\rm DM}(q) |^2 \eta(E_{ \rm min}),
\end{align}
where the DM form factor $F_{\rm DM}$ can be taken to 1 if the interaction is short range. 
We only consider cases where
the angular dependence is trivial, $q = |\vec q|$. The dimensionless
atomic form factor describing the strength of the ionization process
from atomic state $n,l$ is given by
\begin{align}\nonumber
  |f_{nl}({q, p_e'})|^2 = \frac{p_e'}{\pi^2 q} \int_{|p_e'-q|}^{p_e'+q} dp' \, p' \sum_{m=-l}^l
  | \langle \vec p_e' | e^{ i \vec q\cdot \vec r} | nlm
  \rangle |^2 .
\end{align}
We evaluate the latter using radial Hartree-Fock atomic wavefunctions
$R_{nl}(r)$~\cite{BUNGE1993113} in
$\psi_{nlm}(\vec r) = R_{nl}(r)Y_{lm}(\hat r)$ and the plane wave
approximation $|\vec p_e'\rangle = e^{i \vec p_e' \cdot \vec r }$,
including a Sommerfeld factor with effective charge
$Z_{\rm eff} = 1$~\cite{Essig:2011nj}; $p_e' = \sqrt{2 m_e E_{R,e}}$.
When $m_{\rm DM}\ll 0.1\,\rm MeV$, $\vec q \cdot \vec r \ll 1$ is
possible. In order to avoid spurious contributions to $f_{nl}$
from potential numerical non-orthogonality in
$ \langle \vec p_e' | 1 |nlm \rangle $, we subtract the identity
operator, and evaluate
$ \langle \vec p_e' | e^{ i \vec q\cdot \vec r} - 1 | nlm \rangle $ in
these cases instead.
The event rate from level $(n,\ l)$ is then determined by evaluating the
average over the incoming energy spectrum of the reflected DM
component, that in the nonrelativistic limit is
$ \eta(E_{ \rm min}) = \int_{E_{\rm min}} dE (m_{\rm DM}/(2E))^{1/2} (d\Phi_{\rm refl}/dE)\Phi_{\rm halo}^{-1}$.
Multiplying it by the flux and target density $N_T$, we arrive at the total rate from the $(n,l)$ state,
$dR_{nl}/d\ln E_{R,e} = N_T \Phi_{\rm halo} d \langle \sigma_{nl}
v \rangle / d\ln E_{R,e} $, where
$E_{ \rm min}$ is the minimum DM energy required to produce an electron
with $E_{R,e}$ recoil energy.

The resulting electron recoil energy spectrum is converted into
scintillation (S1) and ionization (S2) responses in liquid xenon
experiments,
$dR_{nl}/d\Si = \varepsilon(\Si) \int dE_{R,e}\, \pdf(\Si|E^{nl}_{\rm
  dep.})  dR/d\ln E_{R,e} $.
Here, $\varepsilon(\Si)$ is the $\Si$ detection efficiency and
$\pdf(\Si|E_{\rm dep.})$ is the probability to produce $\Si$ given a
deposited energy $E^{nl}_{\rm dep.} = E_{R,e} + |E_{B}^{nl}|$. For the
purpose of this work, we consider the signals in \Sone\ and \Stwo\
separately,
and model
$\pdf(\Si|E^{nl}_{\rm dep.})$ as follows: the number of
produced quanta at the interaction point is
$ N_Q = {E_{\rm dep.}}/{W} $ with
$W = 13.7\,\eV$~\cite{Dahl:2009nta,Akerib:2016qlr}, partitioned into
$n_e$ ionized electrons escaping the interaction point and $n_\gamma$
scintillation photons.
The latter follow a binomial distribution with $N_Q$ trials and single
event probability $f_{e,\gamma} = \langle n_{e,\gamma} \rangle /N_Q$.
For the purpose of setting limits we only use data above
$E_{\rm dep.} = 0.19\,\keV$ for computing $ \langle n_e \rangle$,
corresponding to the lowest measured charge yield~\cite{Xe127}
(together with \cite{Akerib:2015wdi}; see also
\cite{Szydagis:2013sih,Goetzke:2016lfg}), and determine the light output
self-consistently by demanding conservation of $N_Q$.

\begin{figure}
 \centerline{
 \includegraphics[width=\columnwidth]{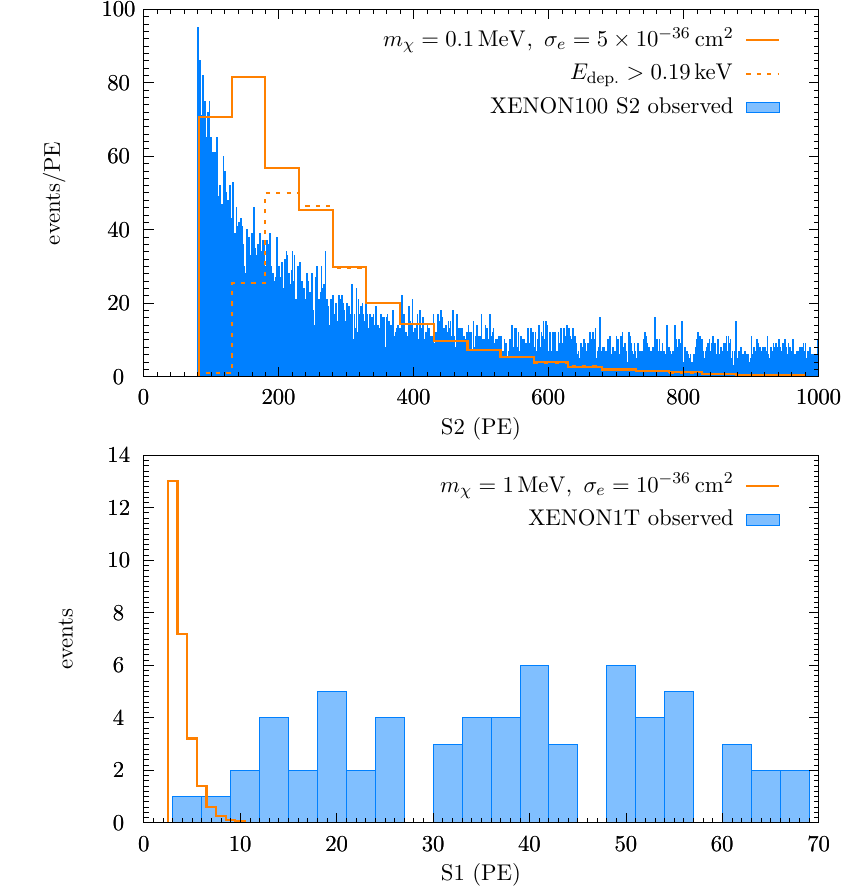}
}
\caption{\footnotesize Exemplary electron scattering event rates as a function
  of S2 in XENON100 (upper panel) and as a function of S1 for XENON1T
  (lower panel). When setting limits we require a minimum deposited
  energy of $E_{\rm dep.}>0.19\ \keV$ (dashed curve).}
 \label{S1S2}
\end{figure}

The detected signals are related by $N_Q = S1/g_1 + S2/g_2$ where
$g_1$ is the light collection efficiency and $g_2$ is the electron
scintillation response times the  electron extraction
efficiency at the gas-liquid interface. 
For computing S1 we use the respective values
$g_1 = 0.12,\ 0.1134,\ 0.144,\ 0.1\ \PE/\gamma$ for
XENON100~\cite{Aprile:2014eoa}, PandaX-II (run 10)~\cite{Cui:2017nnn},
XENON1T~\cite{Aprile:2017iyp}, and LZ~\cite{Mount:2017qzi}. For
computing S2 we use the respective values $g_2 = 20,\ 12.1\ \PE/e^-$
for XENON100~\cite{Aprile:2016wwo} and LUX~\cite{Akerib:2017uem}; for
XENON10, the data has already been converted from S2 to the number of
electrons~\cite{Angle:2011th}. S1 is sampled from a binomial
distribution with $n_{\gamma}$ trials and detection probability $g_1$;
a Gaussian PMT resolution of
$\sigma_{\rm PMT}/\sqrt{\tilde n_{\gamma}} = 0.4\ \PE$ in detected
photons $\tilde n_{\gamma}$ is included.
For S2 we assume an average 80\% electron drift survival probability
and apply a representative Gaussian width of
$\sigma_{\Stwo}/\sqrt{\tilde n_e} = 7\ \PE$~\cite{Aprile:2013blg} in
the conversion of successfully drifted electrons, $\tilde n_e$, to~S2.
After accounting for detection efficiencies, and respecting the
nominal thresholds in the various experiments, the generated signals
are compared to data 
as reported
in~\cite{Aprile:2014eoa,Akerib:2017uem,Tan:2016zwf,Aprile:2017iyp}
and~\cite{Angle:2011th,S1effXe1T,Aprile:2016wwo} for S1 and S2-only,
respectively.  Exemplary spectra for S2 in XENON100 and for S1 in
XENON1T are shown in Fig.~\ref{S1S2}. In the final step, we use the
`$p_{\rm max}$ method'~\cite{Angle:2011th,Yellin:2002xd} to arrive at
the limits in the plane of $ \sigma_e$ and~$m_{\rm DM}$.
 
To complete this analysis, we highlight the principal reach of future
direct detection experiments (making optimistic assumptions.)
For LZ, the next generation liquid
xenon experiment~\cite{Mount:2017qzi}, we assume, for simplicity,
100\% detection efficiency in the acceptance region
$\Sone \geq 3\, \PE$ and include the solar neutrino generated
background in the electron recoil band~\cite{McCabe:2017rln}. For
future semiconductor experiments, we employ the ionization form factor
computed in~\cite{Essig:2015cda} and apply it to a straightforward
generalization of~(\ref{eq:ddrate}); we then follow the
recommendations of~\cite{Essig:2015cda} to obtain the projections for
SENSEI~\cite{Tiffenberg:2017aac} (superCDMS~\cite{Agnese:2016cpb}) with
100~g-yr (10~kg-yr) background-free exposure and 2$e^-$ (1$e^-$)
ionization threshold. The results are summarized in
Fig.~\ref{results}.
Further details are found in a supplement.

\paragraph{Constraints on Light DM Models.}

To demonstrate the application of our analysis, we consider a complex scalar dark matter candidate interacting with the electron vector current, 
\begin{equation}
  \label{eq:model}
{\cal L}_{\rm int} = G_{\chi e} \times (\bar e \gamma^{\mu} e) (i\chi^*\partial_\mu \chi - i \chi \partial_\mu \chi^*).
\end{equation}
This model has been analyzed thoroughly, in particular when the interaction is rendered UV-complete via introduction of a 
kinetically mixed `dark photon' \cite{Holdom:1985ag,Pospelov:2007mp,deNiverville:2011it}. 
The $p$-wave annihilation channel allows this model to escape stringent CMB constraints~\cite{Slatyer:2015jla}.
Carrying out the standard freeze-out calculation, and adjusting the coupling in $\langle \sigma_{\rm ann}v_{\rm rel} \rangle$ 
to reproduce the correct relic abundance as a function of $m_\chi$, we arrive at the  scattering cross section given by
\begin{equation*}
\sigma_e = \frac{1}{\pi}G_{\chi e}^2 \mu^2_{\ch,e}~ \to ~ (8{-}9)\times 10^{-35}\,{\rm cm^2} \, \times  \,  \frac{2\mu^2_{\ch,e} }{(2m_\chi^2+m_e^2)v_e},
\end{equation*}
where $v_e^2= 1- m_e^2/m_\chi^2$.  When $m_\chi$ is close to or below
$m_e$, a more accurate thermal average is required, which we implement
numerically following Refs.~\cite{Gondolo:1990dk,Belanger:2006is}. The
resulting contour is plotted in Fig.~\ref{results}, and one observes
that the reflected DM scattering analysis excludes $m_\chi <2\,$MeV
region, while higher masses are currently allowed.

 Going further afield in `model space', 
 there is now an increased focus on
variants of the thermal relic (or WIMP) paradigm, that can ensure the
correct relic abundance over the MeV mass range, \textit{e.g.}, SIMPs
\cite{Hochberg:2014dra}, ELDERS \cite{Kuflik:2015isi}, and models
utilizing freeze-in production with very light mediators (so that
$F_{\rm DM}(q) = (\al m_e/q)^2$). The latter case is of interest, as
the target parameters correspond to
$\bar{\si}_e \sim 10^{-37}-10^{-38}$ cm$^2$, for
$m_\ch \sim 100-1000$~keV \cite{Essig:2015cda}, which provide a
challenging goal for future experiments.

\paragraph{ Discussion.} 

We have analyzed the direct detection sensitivity to DM-electron scattering, via an energetic `reflected DM' flux produced through re-scattering in the Sun. This leads to new sensitivity at the sub-pb level for light dark matter in the sub-MeV mass range. Similar re-scattering can also occur within the Earth, which would be of particular interest in producing daily modulation. However, the up-scattering effect would be less significant due to the lower electron temperature.  

The limits shown in Fig.~2 apply to all DM models with significant
scattering cross sections on electrons. However, models in this mass
range are subject to a number of 
powerful indirect constraints. 
Besides the CMB-anisotropy-derived limits on annhilation of DM, there
are constraints from stellar energy loss, and the measured radiation
energy density, $N_{\rm eff}$, as well as from primordial
nucleosynthesis (BBN)
\cite{Nollett:2013pwa,Nollett:2014lwa,Boehm:2013jpa}.  A universally
safe way of escaping the BBN and $N_{\rm eff}$ bounds is to consider
$m_{\chi} >$ few MeV.  Internally thermalized DM models with a lower
mass can avoid the constraint on $N_{\rm eff}$ (which in these
models is generally shifted below 3), by annihilating into a mixture of SM
states (\textit{e.g.}~photons) and neutrino-like dark radiation, as
there are compensating effects on the number of equivalent neutrinos
\cite{Nollett:2013pwa,Nollett:2014lwa}. We emphasize that the new
constraints derived on $\sigma_e$ in this paper are direct, and
largely independent of additional particle content in the early
universe.

We conclude by emphasizing that `reflected DM' is an intrinsic
contribution to the DM flux, and can be probed by all upcoming
experiments with sensitivity to electron scattering, e.g. SENSEI,
CRESST-III~\cite{Strauss:2016sxp}, SuperCDMS, LZ, and CDEX-1T~\cite{Kang:2013sjq}.
We leave a study of other DM models as well as an investigation of potential signal/background discrimination for future work~\cite{future}.

\paragraph{ Acknowledgements.} We thank T.~Emken and N.~G.~Nielsen for
pointing out the importance of finite electron velocity inside the
sun.  HA is supported by the Walter Burke Institute at Caltech and by
DOE Grant DE-SC0011632. The work of MP and AR is supported in part by
NSERC, Canada, and research at the Perimeter Institute is supported in
part by the Government of Canada through NSERC and by the Province of
Ontario through MEDT. JP is supported by the New Frontiers program of
the Austrian Academy of Sciences.

\bibliography{Ref}

\begin{thebibliography}{67}%
\makeatletter
\providecommand \@ifxundefined [1]{%
 \@ifx{#1\undefined}
}%
\providecommand \@ifnum [1]{%
 \ifnum #1\expandafter \@firstoftwo
 \else \expandafter \@secondoftwo
 \fi
}%
\providecommand \@ifx [1]{%
 \ifx #1\expandafter \@firstoftwo
 \else \expandafter \@secondoftwo
 \fi
}%
\providecommand \natexlab [1]{#1}%
\providecommand \enquote  [1]{``#1''}%
\providecommand \bibnamefont  [1]{#1}%
\providecommand \bibfnamefont [1]{#1}%
\providecommand \citenamefont [1]{#1}%
\providecommand \href@noop [0]{\@secondoftwo}%
\providecommand \href [0]{\begingroup \@sanitize@url \@href}%
\providecommand \@href[1]{\@@startlink{#1}\@@href}%
\providecommand \@@href[1]{\endgroup#1\@@endlink}%
\providecommand \@sanitize@url [0]{\catcode `\\12\catcode `\$12\catcode
  `\&12\catcode `\#12\catcode `\^12\catcode `\_12\catcode `\%12\relax}%
\providecommand \@@startlink[1]{}%
\providecommand \@@endlink[0]{}%
\providecommand \url  [0]{\begingroup\@sanitize@url \@url }%
\providecommand \@url [1]{\endgroup\@href {#1}{\urlprefix }}%
\providecommand \urlprefix  [0]{URL }%
\providecommand \Eprint [0]{\href }%
\providecommand \doibase [0]{http://dx.doi.org/}%
\providecommand \selectlanguage [0]{\@gobble}%
\providecommand \bibinfo  [0]{\@secondoftwo}%
\providecommand \bibfield  [0]{\@secondoftwo}%
\providecommand \translation [1]{[#1]}%
\providecommand \BibitemOpen [0]{}%
\providecommand \bibitemStop [0]{}%
\providecommand \bibitemNoStop [0]{.\EOS\space}%
\providecommand \EOS [0]{\spacefactor3000\relax}%
\providecommand \BibitemShut  [1]{\csname bibitem#1\endcsname}%
\let\auto@bib@innerbib\@empty
\bibitem [{\citenamefont {Aprile}\ \emph {et~al.}(2017)\citenamefont {Aprile}
  \emph {et~al.}}]{Aprile:2017iyp}%
  \BibitemOpen
  \bibfield  {author} {\bibinfo {author} {\bibfnamefont {E.}~\bibnamefont
  {Aprile}} \emph {et~al.} (\bibinfo {collaboration} {XENON}),\ }\href@noop {}
  {\  (\bibinfo {year} {2017})},\ \Eprint {http://arxiv.org/abs/1705.06655}
  {arXiv:1705.06655 [astro-ph.CO]} \BibitemShut {NoStop}%
\bibitem [{\citenamefont {Akerib}\ \emph
  {et~al.}(2017{\natexlab{a}})\citenamefont {Akerib} \emph
  {et~al.}}]{Akerib:2016vxi}%
  \BibitemOpen
  \bibfield  {author} {\bibinfo {author} {\bibfnamefont {D.~S.}\ \bibnamefont
  {Akerib}} \emph {et~al.} (\bibinfo {collaboration} {LUX}),\ }\href {\doibase
  10.1103/PhysRevLett.118.021303} {\bibfield  {journal} {\bibinfo  {journal}
  {Phys. Rev. Lett.}\ }\textbf {\bibinfo {volume} {118}},\ \bibinfo {pages}
  {021303} (\bibinfo {year} {2017}{\natexlab{a}})},\ \Eprint
  {http://arxiv.org/abs/1608.07648} {arXiv:1608.07648 [astro-ph.CO]}
  \BibitemShut {NoStop}%
\bibitem [{\citenamefont {Tan}\ \emph {et~al.}(2016)\citenamefont {Tan} \emph
  {et~al.}}]{Tan:2016zwf}%
  \BibitemOpen
  \bibfield  {author} {\bibinfo {author} {\bibfnamefont {A.}~\bibnamefont
  {Tan}} \emph {et~al.} (\bibinfo {collaboration} {PandaX-II}),\ }\href
  {\doibase 10.1103/PhysRevLett.117.121303} {\bibfield  {journal} {\bibinfo
  {journal} {Phys. Rev. Lett.}\ }\textbf {\bibinfo {volume} {117}},\ \bibinfo
  {pages} {121303} (\bibinfo {year} {2016})},\ \Eprint
  {http://arxiv.org/abs/1607.07400} {arXiv:1607.07400 [hep-ex]} \BibitemShut
  {NoStop}%
\bibitem [{\citenamefont {Angloher}\ \emph {et~al.}(2016)\citenamefont
  {Angloher} \emph {et~al.}}]{Angloher:2015ewa}%
  \BibitemOpen
  \bibfield  {author} {\bibinfo {author} {\bibfnamefont {G.}~\bibnamefont
  {Angloher}} \emph {et~al.} (\bibinfo {collaboration} {CRESST}),\ }\href
  {\doibase 10.1140/epjc/s10052-016-3877-3} {\bibfield  {journal} {\bibinfo
  {journal} {Eur. Phys. J.}\ }\textbf {\bibinfo {volume} {C76}},\ \bibinfo
  {pages} {25} (\bibinfo {year} {2016})},\ \Eprint
  {http://arxiv.org/abs/1509.01515} {arXiv:1509.01515 [astro-ph.CO]}
  \BibitemShut {NoStop}%
\bibitem [{\citenamefont {Kouvaris}\ and\ \citenamefont
  {Pradler}(2017)}]{Kouvaris:2016afs}%
  \BibitemOpen
  \bibfield  {author} {\bibinfo {author} {\bibfnamefont {C.}~\bibnamefont
  {Kouvaris}}\ and\ \bibinfo {author} {\bibfnamefont {J.}~\bibnamefont
  {Pradler}},\ }\href {\doibase 10.1103/PhysRevLett.118.031803} {\bibfield
  {journal} {\bibinfo  {journal} {Phys. Rev. Lett.}\ }\textbf {\bibinfo
  {volume} {118}},\ \bibinfo {pages} {031803} (\bibinfo {year} {2017})},\
  \Eprint {http://arxiv.org/abs/1607.01789} {arXiv:1607.01789 [hep-ph]}
  \BibitemShut {NoStop}%
\bibitem [{\citenamefont {McCabe}(2017)}]{McCabe:2017rln}%
  \BibitemOpen
  \bibfield  {author} {\bibinfo {author} {\bibfnamefont {C.}~\bibnamefont
  {McCabe}},\ }\href@noop {} {\  (\bibinfo {year} {2017})},\ \Eprint
  {http://arxiv.org/abs/1702.04730} {arXiv:1702.04730 [hep-ph]} \BibitemShut
  {NoStop}%
\bibitem [{\citenamefont {Battaglieri}\ \emph {et~al.}(2017)\citenamefont
  {Battaglieri} \emph {et~al.}}]{Battaglieri:2017aum}%
  \BibitemOpen
  \bibfield  {author} {\bibinfo {author} {\bibfnamefont {M.}~\bibnamefont
  {Battaglieri}} \emph {et~al.},\ }\href@noop {} {\  (\bibinfo {year}
  {2017})},\ \Eprint {http://arxiv.org/abs/1707.04591} {arXiv:1707.04591
  [hep-ph]} \BibitemShut {NoStop}%
\bibitem [{\citenamefont {Alexander}\ \emph {et~al.}(2016)\citenamefont
  {Alexander} \emph {et~al.}}]{Alexander:2016aln}%
  \BibitemOpen
  \bibfield  {author} {\bibinfo {author} {\bibfnamefont {J.}~\bibnamefont
  {Alexander}} \emph {et~al.}\ }(\bibinfo {year} {2016})\ \Eprint
  {http://arxiv.org/abs/1608.08632} {arXiv:1608.08632 [hep-ph]} \BibitemShut
  {NoStop}%
\bibitem [{\citenamefont {Boehm}\ and\ \citenamefont
  {Fayet}(2004)}]{Boehm:2003hm}%
  \BibitemOpen
  \bibfield  {author} {\bibinfo {author} {\bibfnamefont {C.}~\bibnamefont
  {Boehm}}\ and\ \bibinfo {author} {\bibfnamefont {P.}~\bibnamefont {Fayet}},\
  }\href {\doibase 10.1016/j.nuclphysb.2004.01.015} {\bibfield  {journal}
  {\bibinfo  {journal} {Nucl.Phys.}\ }\textbf {\bibinfo {volume} {B683}},\
  \bibinfo {pages} {219} (\bibinfo {year} {2004})},\ \Eprint
  {http://arxiv.org/abs/hep-ph/0305261} {arXiv:hep-ph/0305261 [hep-ph]}
  \BibitemShut {NoStop}%
\bibitem [{\citenamefont {Batell}\ \emph {et~al.}(2009)\citenamefont {Batell},
  \citenamefont {Pospelov},\ and\ \citenamefont {Ritz}}]{Batell:2009di}%
  \BibitemOpen
  \bibfield  {author} {\bibinfo {author} {\bibfnamefont {B.}~\bibnamefont
  {Batell}}, \bibinfo {author} {\bibfnamefont {M.}~\bibnamefont {Pospelov}}, \
  and\ \bibinfo {author} {\bibfnamefont {A.}~\bibnamefont {Ritz}},\ }\href
  {\doibase 10.1103/PhysRevD.80.095024} {\bibfield  {journal} {\bibinfo
  {journal} {Phys.Rev.}\ }\textbf {\bibinfo {volume} {D80}},\ \bibinfo {pages}
  {095024} (\bibinfo {year} {2009})},\ \Eprint {http://arxiv.org/abs/0906.5614}
  {arXiv:0906.5614 [hep-ph]} \BibitemShut {NoStop}%
\bibitem [{\citenamefont {deNiverville}\ \emph {et~al.}(2011)\citenamefont
  {deNiverville}, \citenamefont {Pospelov},\ and\ \citenamefont
  {Ritz}}]{deNiverville:2011it}%
  \BibitemOpen
  \bibfield  {author} {\bibinfo {author} {\bibfnamefont {P.}~\bibnamefont
  {deNiverville}}, \bibinfo {author} {\bibfnamefont {M.}~\bibnamefont
  {Pospelov}}, \ and\ \bibinfo {author} {\bibfnamefont {A.}~\bibnamefont
  {Ritz}},\ }\href {\doibase 10.1103/PhysRevD.84.075020} {\bibfield  {journal}
  {\bibinfo  {journal} {Phys.Rev.}\ }\textbf {\bibinfo {volume} {D84}},\
  \bibinfo {pages} {075020} (\bibinfo {year} {2011})},\ \Eprint
  {http://arxiv.org/abs/1107.4580} {arXiv:1107.4580 [hep-ph]} \BibitemShut
  {NoStop}%
\bibitem [{\citenamefont {deNiverville}\ \emph {et~al.}(2012)\citenamefont
  {deNiverville}, \citenamefont {McKeen},\ and\ \citenamefont
  {Ritz}}]{deNiverville:2012ij}%
  \BibitemOpen
  \bibfield  {author} {\bibinfo {author} {\bibfnamefont {P.}~\bibnamefont
  {deNiverville}}, \bibinfo {author} {\bibfnamefont {D.}~\bibnamefont
  {McKeen}}, \ and\ \bibinfo {author} {\bibfnamefont {A.}~\bibnamefont
  {Ritz}},\ }\href {\doibase 10.1103/PhysRevD.86.035022} {\bibfield  {journal}
  {\bibinfo  {journal} {Phys.Rev.}\ }\textbf {\bibinfo {volume} {D86}},\
  \bibinfo {pages} {035022} (\bibinfo {year} {2012})},\ \Eprint
  {http://arxiv.org/abs/1205.3499} {arXiv:1205.3499 [hep-ph]} \BibitemShut
  {NoStop}%
\bibitem [{\citenamefont {Bjorken}\ \emph {et~al.}(2009)\citenamefont
  {Bjorken}, \citenamefont {Essig}, \citenamefont {Schuster},\ and\
  \citenamefont {Toro}}]{Bjorken:2009mm}%
  \BibitemOpen
  \bibfield  {author} {\bibinfo {author} {\bibfnamefont {J.~D.}\ \bibnamefont
  {Bjorken}}, \bibinfo {author} {\bibfnamefont {R.}~\bibnamefont {Essig}},
  \bibinfo {author} {\bibfnamefont {P.}~\bibnamefont {Schuster}}, \ and\
  \bibinfo {author} {\bibfnamefont {N.}~\bibnamefont {Toro}},\ }\href {\doibase
  10.1103/PhysRevD.80.075018} {\bibfield  {journal} {\bibinfo  {journal}
  {Phys.Rev.}\ }\textbf {\bibinfo {volume} {D80}},\ \bibinfo {pages} {075018}
  (\bibinfo {year} {2009})},\ \Eprint {http://arxiv.org/abs/0906.0580}
  {arXiv:0906.0580 [hep-ph]} \BibitemShut {NoStop}%
\bibitem [{\citenamefont {Kahn}\ \emph {et~al.}(2014)\citenamefont {Kahn},
  \citenamefont {Krnjaic}, \citenamefont {Thaler},\ and\ \citenamefont
  {Toups}}]{Kahn:2014sra}%
  \BibitemOpen
  \bibfield  {author} {\bibinfo {author} {\bibfnamefont {Y.}~\bibnamefont
  {Kahn}}, \bibinfo {author} {\bibfnamefont {G.}~\bibnamefont {Krnjaic}},
  \bibinfo {author} {\bibfnamefont {J.}~\bibnamefont {Thaler}}, \ and\ \bibinfo
  {author} {\bibfnamefont {M.}~\bibnamefont {Toups}},\ }\href@noop {} {\
  (\bibinfo {year} {2014})},\ \Eprint {http://arxiv.org/abs/1411.1055}
  {arXiv:1411.1055 [hep-ph]} \BibitemShut {NoStop}%
\bibitem [{\citenamefont {Dobrescu}\ and\ \citenamefont
  {Frugiuele}(2015)}]{Dobrescu:2014ita}%
  \BibitemOpen
  \bibfield  {author} {\bibinfo {author} {\bibfnamefont {B.~A.}\ \bibnamefont
  {Dobrescu}}\ and\ \bibinfo {author} {\bibfnamefont {C.}~\bibnamefont
  {Frugiuele}},\ }\href {\doibase 10.1007/JHEP02(2015)019} {\bibfield
  {journal} {\bibinfo  {journal} {JHEP}\ }\textbf {\bibinfo {volume} {02}},\
  \bibinfo {pages} {019} (\bibinfo {year} {2015})},\ \Eprint
  {http://arxiv.org/abs/1410.1566} {arXiv:1410.1566 [hep-ph]} \BibitemShut
  {NoStop}%
\bibitem [{\citenamefont {Izaguirre}\ \emph {et~al.}(2013)\citenamefont
  {Izaguirre}, \citenamefont {Krnjaic}, \citenamefont {Schuster},\ and\
  \citenamefont {Toro}}]{Izaguirre:2013uxa}%
  \BibitemOpen
  \bibfield  {author} {\bibinfo {author} {\bibfnamefont {E.}~\bibnamefont
  {Izaguirre}}, \bibinfo {author} {\bibfnamefont {G.}~\bibnamefont {Krnjaic}},
  \bibinfo {author} {\bibfnamefont {P.}~\bibnamefont {Schuster}}, \ and\
  \bibinfo {author} {\bibfnamefont {N.}~\bibnamefont {Toro}},\ }\href {\doibase
  10.1103/PhysRevD.88.114015} {\bibfield  {journal} {\bibinfo  {journal}
  {Phys.Rev.}\ }\textbf {\bibinfo {volume} {D88}},\ \bibinfo {pages} {114015}
  (\bibinfo {year} {2013})},\ \Eprint {http://arxiv.org/abs/1307.6554}
  {arXiv:1307.6554 [hep-ph]} \BibitemShut {NoStop}%
\bibitem [{\citenamefont {Izaguirre}\ \emph {et~al.}(2014)\citenamefont
  {Izaguirre}, \citenamefont {Krnjaic}, \citenamefont {Schuster},\ and\
  \citenamefont {Toro}}]{Izaguirre:2014dua}%
  \BibitemOpen
  \bibfield  {author} {\bibinfo {author} {\bibfnamefont {E.}~\bibnamefont
  {Izaguirre}}, \bibinfo {author} {\bibfnamefont {G.}~\bibnamefont {Krnjaic}},
  \bibinfo {author} {\bibfnamefont {P.}~\bibnamefont {Schuster}}, \ and\
  \bibinfo {author} {\bibfnamefont {N.}~\bibnamefont {Toro}},\ }\href@noop {}
  {\  (\bibinfo {year} {2014})},\ \Eprint {http://arxiv.org/abs/1403.6826}
  {arXiv:1403.6826 [hep-ph]} \BibitemShut {NoStop}%
\bibitem [{\citenamefont {Batell}\ \emph {et~al.}(2014)\citenamefont {Batell},
  \citenamefont {Essig},\ and\ \citenamefont {Surujon}}]{Batell:2014mga}%
  \BibitemOpen
  \bibfield  {author} {\bibinfo {author} {\bibfnamefont {B.}~\bibnamefont
  {Batell}}, \bibinfo {author} {\bibfnamefont {R.}~\bibnamefont {Essig}}, \
  and\ \bibinfo {author} {\bibfnamefont {Z.}~\bibnamefont {Surujon}},\ }\href
  {\doibase 10.1103/PhysRevLett.113.171802} {\bibfield  {journal} {\bibinfo
  {journal} {Phys.Rev.Lett.}\ }\textbf {\bibinfo {volume} {113}},\ \bibinfo
  {pages} {171802} (\bibinfo {year} {2014})},\ \Eprint
  {http://arxiv.org/abs/1406.2698} {arXiv:1406.2698 [hep-ph]} \BibitemShut
  {NoStop}%
\bibitem [{\citenamefont {Essig}\ \emph
  {et~al.}(2012{\natexlab{a}})\citenamefont {Essig}, \citenamefont {Mardon},\
  and\ \citenamefont {Volansky}}]{Essig:2011nj}%
  \BibitemOpen
  \bibfield  {author} {\bibinfo {author} {\bibfnamefont {R.}~\bibnamefont
  {Essig}}, \bibinfo {author} {\bibfnamefont {J.}~\bibnamefont {Mardon}}, \
  and\ \bibinfo {author} {\bibfnamefont {T.}~\bibnamefont {Volansky}},\ }\href
  {\doibase 10.1103/PhysRevD.85.076007} {\bibfield  {journal} {\bibinfo
  {journal} {Phys. Rev.}\ }\textbf {\bibinfo {volume} {D85}},\ \bibinfo {pages}
  {076007} (\bibinfo {year} {2012}{\natexlab{a}})},\ \Eprint
  {http://arxiv.org/abs/1108.5383} {arXiv:1108.5383 [hep-ph]} \BibitemShut
  {NoStop}%
\bibitem [{\citenamefont {Essig}\ \emph
  {et~al.}(2012{\natexlab{b}})\citenamefont {Essig}, \citenamefont
  {Manalaysay}, \citenamefont {Mardon}, \citenamefont {Sorensen},\ and\
  \citenamefont {Volansky}}]{Essig:2012yx}%
  \BibitemOpen
  \bibfield  {author} {\bibinfo {author} {\bibfnamefont {R.}~\bibnamefont
  {Essig}}, \bibinfo {author} {\bibfnamefont {A.}~\bibnamefont {Manalaysay}},
  \bibinfo {author} {\bibfnamefont {J.}~\bibnamefont {Mardon}}, \bibinfo
  {author} {\bibfnamefont {P.}~\bibnamefont {Sorensen}}, \ and\ \bibinfo
  {author} {\bibfnamefont {T.}~\bibnamefont {Volansky}},\ }\href {\doibase
  10.1103/PhysRevLett.109.021301} {\bibfield  {journal} {\bibinfo  {journal}
  {Phys. Rev. Lett.}\ }\textbf {\bibinfo {volume} {109}},\ \bibinfo {pages}
  {021301} (\bibinfo {year} {2012}{\natexlab{b}})},\ \Eprint
  {http://arxiv.org/abs/1206.2644} {arXiv:1206.2644 [astro-ph.CO]} \BibitemShut
  {NoStop}%
\bibitem [{\citenamefont {Essig}\ \emph {et~al.}(2016)\citenamefont {Essig},
  \citenamefont {Fernandez-Serra}, \citenamefont {Mardon}, \citenamefont
  {Soto}, \citenamefont {Volansky},\ and\ \citenamefont {Yu}}]{Essig:2015cda}%
  \BibitemOpen
  \bibfield  {author} {\bibinfo {author} {\bibfnamefont {R.}~\bibnamefont
  {Essig}}, \bibinfo {author} {\bibfnamefont {M.}~\bibnamefont
  {Fernandez-Serra}}, \bibinfo {author} {\bibfnamefont {J.}~\bibnamefont
  {Mardon}}, \bibinfo {author} {\bibfnamefont {A.}~\bibnamefont {Soto}},
  \bibinfo {author} {\bibfnamefont {T.}~\bibnamefont {Volansky}}, \ and\
  \bibinfo {author} {\bibfnamefont {T.-T.}\ \bibnamefont {Yu}},\ }\href
  {\doibase 10.1007/JHEP05(2016)046} {\bibfield  {journal} {\bibinfo  {journal}
  {JHEP}\ }\textbf {\bibinfo {volume} {05}},\ \bibinfo {pages} {046} (\bibinfo
  {year} {2016})},\ \Eprint {http://arxiv.org/abs/1509.01598} {arXiv:1509.01598
  [hep-ph]} \BibitemShut {NoStop}%
\bibitem [{\citenamefont {Hochberg}\ \emph {et~al.}(2016)\citenamefont
  {Hochberg}, \citenamefont {Zhao},\ and\ \citenamefont
  {Zurek}}]{Hochberg:2015pha}%
  \BibitemOpen
  \bibfield  {author} {\bibinfo {author} {\bibfnamefont {Y.}~\bibnamefont
  {Hochberg}}, \bibinfo {author} {\bibfnamefont {Y.}~\bibnamefont {Zhao}}, \
  and\ \bibinfo {author} {\bibfnamefont {K.~M.}\ \bibnamefont {Zurek}},\ }\href
  {\doibase 10.1103/PhysRevLett.116.011301} {\bibfield  {journal} {\bibinfo
  {journal} {Phys. Rev. Lett.}\ }\textbf {\bibinfo {volume} {116}},\ \bibinfo
  {pages} {011301} (\bibinfo {year} {2016})},\ \Eprint
  {http://arxiv.org/abs/1504.07237} {arXiv:1504.07237 [hep-ph]} \BibitemShut
  {NoStop}%
\bibitem [{\citenamefont {Essig}\ \emph {et~al.}(2017)\citenamefont {Essig},
  \citenamefont {Volansky},\ and\ \citenamefont {Yu}}]{Essig:2017kqs}%
  \BibitemOpen
  \bibfield  {author} {\bibinfo {author} {\bibfnamefont {R.}~\bibnamefont
  {Essig}}, \bibinfo {author} {\bibfnamefont {T.}~\bibnamefont {Volansky}}, \
  and\ \bibinfo {author} {\bibfnamefont {T.-T.}\ \bibnamefont {Yu}},\
  }\href@noop {} {\  (\bibinfo {year} {2017})},\ \Eprint
  {http://arxiv.org/abs/1703.00910} {arXiv:1703.00910 [hep-ph]} \BibitemShut
  {NoStop}%
\bibitem [{\citenamefont {Kouvaris}(2015)}]{Kouvaris:2015nsa}%
  \BibitemOpen
  \bibfield  {author} {\bibinfo {author} {\bibfnamefont {C.}~\bibnamefont
  {Kouvaris}},\ }\href {\doibase 10.1103/PhysRevD.92.075001} {\bibfield
  {journal} {\bibinfo  {journal} {Phys. Rev.}\ }\textbf {\bibinfo {volume}
  {D92}},\ \bibinfo {pages} {075001} (\bibinfo {year} {2015})},\ \Eprint
  {http://arxiv.org/abs/1506.04316} {arXiv:1506.04316 [hep-ph]} \BibitemShut
  {NoStop}%
\bibitem [{\citenamefont {Garani}\ and\ \citenamefont
  {Palomares-Ruiz}(2017)}]{Garani:2017jcj}%
  \BibitemOpen
  \bibfield  {author} {\bibinfo {author} {\bibfnamefont {R.}~\bibnamefont
  {Garani}}\ and\ \bibinfo {author} {\bibfnamefont {S.}~\bibnamefont
  {Palomares-Ruiz}},\ }\href {\doibase 10.1088/1475-7516/2017/05/007}
  {\bibfield  {journal} {\bibinfo  {journal} {JCAP}\ }\textbf {\bibinfo
  {volume} {1705}},\ \bibinfo {pages} {007} (\bibinfo {year} {2017})},\ \Eprint
  {http://arxiv.org/abs/1702.02768} {arXiv:1702.02768 [hep-ph]} \BibitemShut
  {NoStop}%
\bibitem [{\citenamefont {Bahcall}\ \emph {et~al.}(2005)\citenamefont
  {Bahcall}, \citenamefont {Serenelli},\ and\ \citenamefont
  {Basu}}]{Bahcall:2004pz}%
  \BibitemOpen
  \bibfield  {author} {\bibinfo {author} {\bibfnamefont {J.~N.}\ \bibnamefont
  {Bahcall}}, \bibinfo {author} {\bibfnamefont {A.~M.}\ \bibnamefont
  {Serenelli}}, \ and\ \bibinfo {author} {\bibfnamefont {S.}~\bibnamefont
  {Basu}},\ }\href {\doibase 10.1086/428929} {\bibfield  {journal} {\bibinfo
  {journal} {Astrophys. J.}\ }\textbf {\bibinfo {volume} {621}},\ \bibinfo
  {pages} {L85} (\bibinfo {year} {2005})},\ \Eprint
  {http://arxiv.org/abs/astro-ph/0412440} {arXiv:astro-ph/0412440 [astro-ph]}
  \BibitemShut {NoStop}%
\bibitem [{\citenamefont {Bunge}\ \emph {et~al.}(1993)\citenamefont {Bunge},
  \citenamefont {Barrientos},\ and\ \citenamefont {Bunge}}]{BUNGE1993113}%
  \BibitemOpen
  \bibfield  {author} {\bibinfo {author} {\bibfnamefont {C.}~\bibnamefont
  {Bunge}}, \bibinfo {author} {\bibfnamefont {J.}~\bibnamefont {Barrientos}}, \
  and\ \bibinfo {author} {\bibfnamefont {A.}~\bibnamefont {Bunge}},\ }\href
  {\doibase http://dx.doi.org/10.1006/adnd.1993.1003} {\bibfield  {journal}
  {\bibinfo  {journal} {Atomic Data and Nuclear Data Tables}\ }\textbf
  {\bibinfo {volume} {53}},\ \bibinfo {pages} {113 } (\bibinfo {year}
  {1993})}\BibitemShut {NoStop}%
\bibitem [{\citenamefont {Dahl}(2009)}]{Dahl:2009nta}%
  \BibitemOpen
  \bibfield  {author} {\bibinfo {author} {\bibfnamefont {C.~E.}\ \bibnamefont
  {Dahl}},\ }\emph {\bibinfo {title} {{The physics of background discrimination
  in liquid xenon, and first results from Xenon10 in the hunt for WIMP dark
  matter}}},\ \href
  {https://www.princeton.edu/physics/graduate-program/theses/theses-from-2009/E.Dahlthesis.pdf}
  {Ph.D. thesis},\ \bibinfo  {school} {Princeton U.} (\bibinfo {year}
  {2009})\BibitemShut {NoStop}%
\bibitem [{\citenamefont {Akerib}\ \emph
  {et~al.}(2017{\natexlab{b}})\citenamefont {Akerib} \emph
  {et~al.}}]{Akerib:2016qlr}%
  \BibitemOpen
  \bibfield  {author} {\bibinfo {author} {\bibfnamefont {D.~S.}\ \bibnamefont
  {Akerib}} \emph {et~al.} (\bibinfo {collaboration} {LUX}),\ }\href {\doibase
  10.1103/PhysRevD.95.012008} {\bibfield  {journal} {\bibinfo  {journal} {Phys.
  Rev.}\ }\textbf {\bibinfo {volume} {D95}},\ \bibinfo {pages} {012008}
  (\bibinfo {year} {2017}{\natexlab{b}})},\ \Eprint
  {http://arxiv.org/abs/1610.02076} {arXiv:1610.02076 [physics.ins-det]}
  \BibitemShut {NoStop}%
\bibitem [{\citenamefont {Huang}()}]{Xe127}%
  \BibitemOpen
  \bibfield  {author} {\bibinfo {author} {\bibfnamefont {D.}~\bibnamefont
  {Huang}},\ }\href@noop {} {\ }\Eprint {http://arxiv.org/abs/Ultra-low Energy
  Calibration of LUX Detector using D-D Neutron, Tritium and 127Xe, UCLA DM
  2016} {Ultra-low Energy Calibration of LUX Detector using D-D Neutron,
  Tritium and 127Xe, UCLA DM 2016} \BibitemShut {NoStop}%
\bibitem [{\citenamefont {Akerib}\ \emph {et~al.}(2016)\citenamefont {Akerib}
  \emph {et~al.}}]{Akerib:2015wdi}%
  \BibitemOpen
  \bibfield  {author} {\bibinfo {author} {\bibfnamefont {D.~S.}\ \bibnamefont
  {Akerib}} \emph {et~al.} (\bibinfo {collaboration} {LUX}),\ }\href {\doibase
  10.1103/PhysRevD.93.072009} {\bibfield  {journal} {\bibinfo  {journal} {Phys.
  Rev.}\ }\textbf {\bibinfo {volume} {D93}},\ \bibinfo {pages} {072009}
  (\bibinfo {year} {2016})},\ \Eprint {http://arxiv.org/abs/1512.03133}
  {arXiv:1512.03133 [physics.ins-det]} \BibitemShut {NoStop}%
\bibitem [{\citenamefont {Szydagis}\ \emph {et~al.}(2013)\citenamefont
  {Szydagis}, \citenamefont {Fyhrie}, \citenamefont {Thorngren},\ and\
  \citenamefont {Tripathi}}]{Szydagis:2013sih}%
  \BibitemOpen
  \bibfield  {author} {\bibinfo {author} {\bibfnamefont {M.}~\bibnamefont
  {Szydagis}}, \bibinfo {author} {\bibfnamefont {A.}~\bibnamefont {Fyhrie}},
  \bibinfo {author} {\bibfnamefont {D.}~\bibnamefont {Thorngren}}, \ and\
  \bibinfo {author} {\bibfnamefont {M.}~\bibnamefont {Tripathi}},\ }\bibfield
  {booktitle} {\emph {\bibinfo {booktitle} {{Proceedings, LIght Detection In
  Noble Elements (LIDINE2013): Batavia, USA, May 29-31, 2013}}},\ }\href
  {\doibase 10.1088/1748-0221/8/10/C10003} {\bibfield  {journal} {\bibinfo
  {journal} {JINST}\ }\textbf {\bibinfo {volume} {8}},\ \bibinfo {pages}
  {C10003} (\bibinfo {year} {2013})},\ \Eprint {http://arxiv.org/abs/1307.6601}
  {arXiv:1307.6601 [physics.ins-det]} \BibitemShut {NoStop}%
\bibitem [{\citenamefont {Goetzke}\ \emph {et~al.}(2016)\citenamefont
  {Goetzke}, \citenamefont {Aprile}, \citenamefont {Anthony}, \citenamefont
  {Plante},\ and\ \citenamefont {Weber}}]{Goetzke:2016lfg}%
  \BibitemOpen
  \bibfield  {author} {\bibinfo {author} {\bibfnamefont {L.~W.}\ \bibnamefont
  {Goetzke}}, \bibinfo {author} {\bibfnamefont {E.}~\bibnamefont {Aprile}},
  \bibinfo {author} {\bibfnamefont {M.}~\bibnamefont {Anthony}}, \bibinfo
  {author} {\bibfnamefont {G.}~\bibnamefont {Plante}}, \ and\ \bibinfo {author}
  {\bibfnamefont {M.}~\bibnamefont {Weber}},\ }\href@noop {} {\  (\bibinfo
  {year} {2016})},\ \Eprint {http://arxiv.org/abs/1611.10322} {arXiv:1611.10322
  [astro-ph.IM]} \BibitemShut {NoStop}%
\bibitem [{\citenamefont {Aprile}\ \emph
  {et~al.}(2014{\natexlab{a}})\citenamefont {Aprile} \emph
  {et~al.}}]{Aprile:2014eoa}%
  \BibitemOpen
  \bibfield  {author} {\bibinfo {author} {\bibfnamefont {E.}~\bibnamefont
  {Aprile}} \emph {et~al.} (\bibinfo {collaboration} {XENON100}),\ }\href
  {\doibase 10.1103/PhysRevD.90.062009, 10.1103/PhysRevD.95.029904} {\bibfield
  {journal} {\bibinfo  {journal} {Phys. Rev.}\ }\textbf {\bibinfo {volume}
  {D90}},\ \bibinfo {pages} {062009} (\bibinfo {year} {2014}{\natexlab{a}})},\
  \bibinfo {note} {[Erratum: Phys. Rev.D95,no.2,029904(2017)]},\ \Eprint
  {http://arxiv.org/abs/1404.1455} {arXiv:1404.1455 [astro-ph.CO]} \BibitemShut
  {NoStop}%
\bibitem [{\citenamefont {Cui}\ \emph {et~al.}(2017)\citenamefont {Cui} \emph
  {et~al.}}]{Cui:2017nnn}%
  \BibitemOpen
  \bibfield  {author} {\bibinfo {author} {\bibfnamefont {X.}~\bibnamefont
  {Cui}} \emph {et~al.} (\bibinfo {collaboration} {PandaX-II}),\ }\href
  {\doibase 10.1103/PhysRevLett.119.181302} {\bibfield  {journal} {\bibinfo
  {journal} {Phys. Rev. Lett.}\ }\textbf {\bibinfo {volume} {119}},\ \bibinfo
  {pages} {181302} (\bibinfo {year} {2017})},\ \Eprint
  {http://arxiv.org/abs/1708.06917} {arXiv:1708.06917 [astro-ph.CO]}
  \BibitemShut {NoStop}%
\bibitem [{\citenamefont {Mount}\ \emph {et~al.}(2017)\citenamefont {Mount}
  \emph {et~al.}}]{Mount:2017qzi}%
  \BibitemOpen
  \bibfield  {author} {\bibinfo {author} {\bibfnamefont {B.~J.}\ \bibnamefont
  {Mount}} \emph {et~al.},\ }\href@noop {} {\  (\bibinfo {year} {2017})},\
  \Eprint {http://arxiv.org/abs/1703.09144} {arXiv:1703.09144
  [physics.ins-det]} \BibitemShut {NoStop}%
\bibitem [{\citenamefont {Aprile}\ \emph {et~al.}(2016)\citenamefont {Aprile}
  \emph {et~al.}}]{Aprile:2016wwo}%
  \BibitemOpen
  \bibfield  {author} {\bibinfo {author} {\bibfnamefont {E.}~\bibnamefont
  {Aprile}} \emph {et~al.} (\bibinfo {collaboration} {XENON}),\ }\href
  {\doibase 10.1103/PhysRevD.94.092001, 10.1103/PhysRevD.95.059901} {\bibfield
  {journal} {\bibinfo  {journal} {Phys. Rev.}\ }\textbf {\bibinfo {volume}
  {D94}},\ \bibinfo {pages} {092001} (\bibinfo {year} {2016})},\ \bibinfo
  {note} {[Erratum: Phys. Rev.D95,no.5,059901(2017)]},\ \Eprint
  {http://arxiv.org/abs/1605.06262} {arXiv:1605.06262 [astro-ph.CO]}
  \BibitemShut {NoStop}%
\bibitem [{\citenamefont {Akerib}\ \emph
  {et~al.}(2017{\natexlab{c}})\citenamefont {Akerib} \emph
  {et~al.}}]{Akerib:2017uem}%
  \BibitemOpen
  \bibfield  {author} {\bibinfo {author} {\bibfnamefont {D.~S.}\ \bibnamefont
  {Akerib}} \emph {et~al.} (\bibinfo {collaboration} {LUX}),\ }\href {\doibase
  10.1103/PhysRevLett.118.261301} {\bibfield  {journal} {\bibinfo  {journal}
  {Phys. Rev. Lett.}\ }\textbf {\bibinfo {volume} {118}},\ \bibinfo {pages}
  {261301} (\bibinfo {year} {2017}{\natexlab{c}})},\ \Eprint
  {http://arxiv.org/abs/1704.02297} {arXiv:1704.02297 [astro-ph.CO]}
  \BibitemShut {NoStop}%
\bibitem [{\citenamefont {Angle}\ \emph {et~al.}(2011)\citenamefont {Angle}
  \emph {et~al.}}]{Angle:2011th}%
  \BibitemOpen
  \bibfield  {author} {\bibinfo {author} {\bibfnamefont {J.}~\bibnamefont
  {Angle}} \emph {et~al.} (\bibinfo {collaboration} {XENON10}),\ }\href
  {\doibase 10.1103/PhysRevLett.110.249901, 10.1103/PhysRevLett.107.051301}
  {\bibfield  {journal} {\bibinfo  {journal} {Phys. Rev. Lett.}\ }\textbf
  {\bibinfo {volume} {107}},\ \bibinfo {pages} {051301} (\bibinfo {year}
  {2011})},\ \bibinfo {note} {[Erratum: Phys. Rev. Lett.110,249901(2013)]},\
  \Eprint {http://arxiv.org/abs/1104.3088} {arXiv:1104.3088 [astro-ph.CO]}
  \BibitemShut {NoStop}%
\bibitem [{\citenamefont {Aprile}\ \emph
  {et~al.}(2014{\natexlab{b}})\citenamefont {Aprile} \emph
  {et~al.}}]{Aprile:2013blg}%
  \BibitemOpen
  \bibfield  {author} {\bibinfo {author} {\bibfnamefont {E.}~\bibnamefont
  {Aprile}} \emph {et~al.} (\bibinfo {collaboration} {XENON100}),\ }\href
  {\doibase 10.1088/0954-3899/41/3/035201} {\bibfield  {journal} {\bibinfo
  {journal} {J. Phys.}\ }\textbf {\bibinfo {volume} {G41}},\ \bibinfo {pages}
  {035201} (\bibinfo {year} {2014}{\natexlab{b}})},\ \Eprint
  {http://arxiv.org/abs/1311.1088} {arXiv:1311.1088 [physics.ins-det]}
  \BibitemShut {NoStop}%
\bibitem [{\citenamefont {Aprile}()}]{S1effXe1T}%
  \BibitemOpen
  \bibfield  {author} {\bibinfo {author} {\bibfnamefont {E.}~\bibnamefont
  {Aprile}},\ }\href@noop {} {\ }\Eprint {http://arxiv.org/abs/XENON1T: First
  Results, Patras Axion-Wimp 2017} {XENON1T: First Results, Patras Axion-Wimp
  2017} \BibitemShut {NoStop}%
\bibitem [{\citenamefont {Yellin}(2002)}]{Yellin:2002xd}%
  \BibitemOpen
  \bibfield  {author} {\bibinfo {author} {\bibfnamefont {S.}~\bibnamefont
  {Yellin}},\ }\href {\doibase 10.1103/PhysRevD.66.032005} {\bibfield
  {journal} {\bibinfo  {journal} {Phys. Rev.}\ }\textbf {\bibinfo {volume}
  {D66}},\ \bibinfo {pages} {032005} (\bibinfo {year} {2002})},\ \Eprint
  {http://arxiv.org/abs/physics/0203002} {arXiv:physics/0203002 [physics]}
  \BibitemShut {NoStop}%
\bibitem [{\citenamefont {Tiffenberg}\ \emph {et~al.}(2017)\citenamefont
  {Tiffenberg}, \citenamefont {Sofo-Haro}, \citenamefont {Drlica-Wagner},
  \citenamefont {Essig}, \citenamefont {Guardincerri}, \citenamefont {Holland},
  \citenamefont {Volansky},\ and\ \citenamefont {Yu}}]{Tiffenberg:2017aac}%
  \BibitemOpen
  \bibfield  {author} {\bibinfo {author} {\bibfnamefont {J.}~\bibnamefont
  {Tiffenberg}}, \bibinfo {author} {\bibfnamefont {M.}~\bibnamefont
  {Sofo-Haro}}, \bibinfo {author} {\bibfnamefont {A.}~\bibnamefont
  {Drlica-Wagner}}, \bibinfo {author} {\bibfnamefont {R.}~\bibnamefont
  {Essig}}, \bibinfo {author} {\bibfnamefont {Y.}~\bibnamefont {Guardincerri}},
  \bibinfo {author} {\bibfnamefont {S.}~\bibnamefont {Holland}}, \bibinfo
  {author} {\bibfnamefont {T.}~\bibnamefont {Volansky}}, \ and\ \bibinfo
  {author} {\bibfnamefont {T.-T.}\ \bibnamefont {Yu}},\ }\href {\doibase
  10.1103/PhysRevLett.119.131802} {\bibfield  {journal} {\bibinfo  {journal}
  {Phys. Rev. Lett.}\ }\textbf {\bibinfo {volume} {119}},\ \bibinfo {pages}
  {131802} (\bibinfo {year} {2017})},\ \Eprint
  {http://arxiv.org/abs/1706.00028} {arXiv:1706.00028 [physics.ins-det]}
  \BibitemShut {NoStop}%
\bibitem [{\citenamefont {Agnese}\ \emph {et~al.}(2017)\citenamefont {Agnese}
  \emph {et~al.}}]{Agnese:2016cpb}%
  \BibitemOpen
  \bibfield  {author} {\bibinfo {author} {\bibfnamefont {R.}~\bibnamefont
  {Agnese}} \emph {et~al.} (\bibinfo {collaboration} {SuperCDMS}),\ }\href
  {\doibase 10.1103/PhysRevD.95.082002} {\bibfield  {journal} {\bibinfo
  {journal} {Phys. Rev.}\ }\textbf {\bibinfo {volume} {D95}},\ \bibinfo {pages}
  {082002} (\bibinfo {year} {2017})},\ \Eprint
  {http://arxiv.org/abs/1610.00006} {arXiv:1610.00006 [physics.ins-det]}
  \BibitemShut {NoStop}%
\bibitem [{\citenamefont {Holdom}(1986)}]{Holdom:1985ag}%
  \BibitemOpen
  \bibfield  {author} {\bibinfo {author} {\bibfnamefont {B.}~\bibnamefont
  {Holdom}},\ }\href {\doibase 10.1016/0370-2693(86)91377-8} {\bibfield
  {journal} {\bibinfo  {journal} {Phys.Lett.}\ }\textbf {\bibinfo {volume}
  {B166}},\ \bibinfo {pages} {196} (\bibinfo {year} {1986})}\BibitemShut
  {NoStop}%
\bibitem [{\citenamefont {Pospelov}\ \emph {et~al.}(2008)\citenamefont
  {Pospelov}, \citenamefont {Ritz},\ and\ \citenamefont
  {Voloshin}}]{Pospelov:2007mp}%
  \BibitemOpen
  \bibfield  {author} {\bibinfo {author} {\bibfnamefont {M.}~\bibnamefont
  {Pospelov}}, \bibinfo {author} {\bibfnamefont {A.}~\bibnamefont {Ritz}}, \
  and\ \bibinfo {author} {\bibfnamefont {M.~B.}\ \bibnamefont {Voloshin}},\
  }\href {\doibase 10.1016/j.physletb.2008.02.052} {\bibfield  {journal}
  {\bibinfo  {journal} {Phys.Lett.}\ }\textbf {\bibinfo {volume} {B662}},\
  \bibinfo {pages} {53} (\bibinfo {year} {2008})},\ \Eprint
  {http://arxiv.org/abs/0711.4866} {arXiv:0711.4866 [hep-ph]} \BibitemShut
  {NoStop}%
\bibitem [{\citenamefont {Slatyer}(2016)}]{Slatyer:2015jla}%
  \BibitemOpen
  \bibfield  {author} {\bibinfo {author} {\bibfnamefont {T.~R.}\ \bibnamefont
  {Slatyer}},\ }\href {\doibase 10.1103/PhysRevD.93.023527} {\bibfield
  {journal} {\bibinfo  {journal} {Phys. Rev.}\ }\textbf {\bibinfo {volume}
  {D93}},\ \bibinfo {pages} {023527} (\bibinfo {year} {2016})},\ \Eprint
  {http://arxiv.org/abs/1506.03811} {arXiv:1506.03811 [hep-ph]} \BibitemShut
  {NoStop}%
\bibitem [{\citenamefont {Gondolo}\ and\ \citenamefont
  {Gelmini}(1991)}]{Gondolo:1990dk}%
  \BibitemOpen
  \bibfield  {author} {\bibinfo {author} {\bibfnamefont {P.}~\bibnamefont
  {Gondolo}}\ and\ \bibinfo {author} {\bibfnamefont {G.}~\bibnamefont
  {Gelmini}},\ }\href {\doibase 10.1016/0550-3213(91)90438-4} {\bibfield
  {journal} {\bibinfo  {journal} {Nucl. Phys.}\ }\textbf {\bibinfo {volume}
  {B360}},\ \bibinfo {pages} {145} (\bibinfo {year} {1991})}\BibitemShut
  {NoStop}%
\bibitem [{\citenamefont {Belanger}\ \emph {et~al.}(2007)\citenamefont
  {Belanger}, \citenamefont {Boudjema}, \citenamefont {Pukhov},\ and\
  \citenamefont {Semenov}}]{Belanger:2006is}%
  \BibitemOpen
  \bibfield  {author} {\bibinfo {author} {\bibfnamefont {G.}~\bibnamefont
  {Belanger}}, \bibinfo {author} {\bibfnamefont {F.}~\bibnamefont {Boudjema}},
  \bibinfo {author} {\bibfnamefont {A.}~\bibnamefont {Pukhov}}, \ and\ \bibinfo
  {author} {\bibfnamefont {A.}~\bibnamefont {Semenov}},\ }\href {\doibase
  10.1016/j.cpc.2006.11.008} {\bibfield  {journal} {\bibinfo  {journal}
  {Comput. Phys. Commun.}\ }\textbf {\bibinfo {volume} {176}},\ \bibinfo
  {pages} {367} (\bibinfo {year} {2007})},\ \Eprint
  {http://arxiv.org/abs/hep-ph/0607059} {arXiv:hep-ph/0607059 [hep-ph]}
  \BibitemShut {NoStop}%
\bibitem [{\citenamefont {Hochberg}\ \emph {et~al.}(2014)\citenamefont
  {Hochberg}, \citenamefont {Kuflik}, \citenamefont {Volansky},\ and\
  \citenamefont {Wacker}}]{Hochberg:2014dra}%
  \BibitemOpen
  \bibfield  {author} {\bibinfo {author} {\bibfnamefont {Y.}~\bibnamefont
  {Hochberg}}, \bibinfo {author} {\bibfnamefont {E.}~\bibnamefont {Kuflik}},
  \bibinfo {author} {\bibfnamefont {T.}~\bibnamefont {Volansky}}, \ and\
  \bibinfo {author} {\bibfnamefont {J.~G.}\ \bibnamefont {Wacker}},\ }\href
  {\doibase 10.1103/PhysRevLett.113.171301} {\bibfield  {journal} {\bibinfo
  {journal} {Phys. Rev. Lett.}\ }\textbf {\bibinfo {volume} {113}},\ \bibinfo
  {pages} {171301} (\bibinfo {year} {2014})},\ \Eprint
  {http://arxiv.org/abs/1402.5143} {arXiv:1402.5143 [hep-ph]} \BibitemShut
  {NoStop}%
\bibitem [{\citenamefont {Kuflik}\ \emph {et~al.}(2016)\citenamefont {Kuflik},
  \citenamefont {Perelstein}, \citenamefont {Lorier},\ and\ \citenamefont
  {Tsai}}]{Kuflik:2015isi}%
  \BibitemOpen
  \bibfield  {author} {\bibinfo {author} {\bibfnamefont {E.}~\bibnamefont
  {Kuflik}}, \bibinfo {author} {\bibfnamefont {M.}~\bibnamefont {Perelstein}},
  \bibinfo {author} {\bibfnamefont {N.~R.-L.}\ \bibnamefont {Lorier}}, \ and\
  \bibinfo {author} {\bibfnamefont {Y.-D.}\ \bibnamefont {Tsai}},\ }\href
  {\doibase 10.1103/PhysRevLett.116.221302} {\bibfield  {journal} {\bibinfo
  {journal} {Phys. Rev. Lett.}\ }\textbf {\bibinfo {volume} {116}},\ \bibinfo
  {pages} {221302} (\bibinfo {year} {2016})},\ \Eprint
  {http://arxiv.org/abs/1512.04545} {arXiv:1512.04545 [hep-ph]} \BibitemShut
  {NoStop}%
\bibitem [{\citenamefont {Nollett}\ and\ \citenamefont
  {Steigman}(2014)}]{Nollett:2013pwa}%
  \BibitemOpen
  \bibfield  {author} {\bibinfo {author} {\bibfnamefont {K.~M.}\ \bibnamefont
  {Nollett}}\ and\ \bibinfo {author} {\bibfnamefont {G.}~\bibnamefont
  {Steigman}},\ }\href {\doibase 10.1103/PhysRevD.89.083508} {\bibfield
  {journal} {\bibinfo  {journal} {Phys. Rev.}\ }\textbf {\bibinfo {volume}
  {D89}},\ \bibinfo {pages} {083508} (\bibinfo {year} {2014})},\ \Eprint
  {http://arxiv.org/abs/1312.5725} {arXiv:1312.5725 [astro-ph.CO]} \BibitemShut
  {NoStop}%
\bibitem [{\citenamefont {Nollett}\ and\ \citenamefont
  {Steigman}(2015)}]{Nollett:2014lwa}%
  \BibitemOpen
  \bibfield  {author} {\bibinfo {author} {\bibfnamefont {K.~M.}\ \bibnamefont
  {Nollett}}\ and\ \bibinfo {author} {\bibfnamefont {G.}~\bibnamefont
  {Steigman}},\ }\href {\doibase 10.1103/PhysRevD.91.083505} {\bibfield
  {journal} {\bibinfo  {journal} {Phys. Rev.}\ }\textbf {\bibinfo {volume}
  {D91}},\ \bibinfo {pages} {083505} (\bibinfo {year} {2015})},\ \Eprint
  {http://arxiv.org/abs/1411.6005} {arXiv:1411.6005 [astro-ph.CO]} \BibitemShut
  {NoStop}%
\bibitem [{\citenamefont {Boehm}\ \emph {et~al.}(2013)\citenamefont {Boehm},
  \citenamefont {Dolan},\ and\ \citenamefont {McCabe}}]{Boehm:2013jpa}%
  \BibitemOpen
  \bibfield  {author} {\bibinfo {author} {\bibfnamefont {C.}~\bibnamefont
  {Boehm}}, \bibinfo {author} {\bibfnamefont {M.~J.}\ \bibnamefont {Dolan}}, \
  and\ \bibinfo {author} {\bibfnamefont {C.}~\bibnamefont {McCabe}},\ }\href
  {\doibase 10.1088/1475-7516/2013/08/041} {\bibfield  {journal} {\bibinfo
  {journal} {JCAP}\ }\textbf {\bibinfo {volume} {1308}},\ \bibinfo {pages}
  {041} (\bibinfo {year} {2013})},\ \Eprint {http://arxiv.org/abs/1303.6270}
  {arXiv:1303.6270 [hep-ph]} \BibitemShut {NoStop}%
\bibitem [{\citenamefont {Strauss}\ \emph {et~al.}(2016)\citenamefont {Strauss}
  \emph {et~al.}}]{Strauss:2016sxp}%
  \BibitemOpen
  \bibfield  {author} {\bibinfo {author} {\bibfnamefont {R.}~\bibnamefont
  {Strauss}} \emph {et~al.},\ }\bibfield  {booktitle} {\emph {\bibinfo
  {booktitle} {{Proceedings, 14th International Conference on Topics in
  Astroparticle and Underground Physics (TAUP 2015): Torino, Italy, September
  7-11, 2015}}},\ }\href {\doibase 10.1088/1742-6596/718/4/042048} {\bibfield
  {journal} {\bibinfo  {journal} {J. Phys. Conf. Ser.}\ }\textbf {\bibinfo
  {volume} {718}},\ \bibinfo {pages} {042048} (\bibinfo {year}
  {2016})}\BibitemShut {NoStop}%
\bibitem [{\citenamefont {Kang}\ \emph {et~al.}(2013)\citenamefont {Kang} \emph
  {et~al.}}]{Kang:2013sjq}%
  \BibitemOpen
  \bibfield  {author} {\bibinfo {author} {\bibfnamefont {K.-J.}\ \bibnamefont
  {Kang}} \emph {et~al.} (\bibinfo {collaboration} {CDEX}),\ }\href {\doibase
  10.1007/s11467-013-0349-1} {\bibfield  {journal} {\bibinfo  {journal} {Front.
  Phys.(Beijing)}\ }\textbf {\bibinfo {volume} {8}},\ \bibinfo {pages} {412}
  (\bibinfo {year} {2013})},\ \Eprint {http://arxiv.org/abs/1303.0601}
  {arXiv:1303.0601 [physics.ins-det]} \BibitemShut {NoStop}%
\bibitem [{\citenamefont {An}\ \emph {et~al.}()\citenamefont {An},
  \citenamefont {Pospelov}, \citenamefont {Pradler},\ and\ \citenamefont
  {Ritz}}]{future}%
  \BibitemOpen
  \bibfield  {author} {\bibinfo {author} {\bibfnamefont {H.}~\bibnamefont
  {An}}, \bibinfo {author} {\bibfnamefont {M.}~\bibnamefont {Pospelov}},
  \bibinfo {author} {\bibfnamefont {J.}~\bibnamefont {Pradler}}, \ and\
  \bibinfo {author} {\bibfnamefont {A.}~\bibnamefont {Ritz}},\ }\href@noop {}
  {\ }\Eprint {http://arxiv.org/abs/in preparation} {in preparation}
  \BibitemShut {NoStop}%
\bibitem [{\citenamefont {Akerib}\ \emph
  {et~al.}(2017{\natexlab{d}})\citenamefont {Akerib} \emph
  {et~al.}}]{Akerib:2017hph}%
  \BibitemOpen
  \bibfield  {author} {\bibinfo {author} {\bibfnamefont {D.~S.}\ \bibnamefont
  {Akerib}} \emph {et~al.} (\bibinfo {collaboration} {LUX}),\ }\href@noop {} {\
   (\bibinfo {year} {2017}{\natexlab{d}})},\ \Eprint
  {http://arxiv.org/abs/1709.00800} {arXiv:1709.00800 [physics.ins-det]}
  \BibitemShut {NoStop}%
\bibitem [{\citenamefont {Serenelli}\ \emph {et~al.}(2009)\citenamefont
  {Serenelli}, \citenamefont {Basu}, \citenamefont {Ferguson},\ and\
  \citenamefont {Asplund}}]{Serenelli:2009yc}%
  \BibitemOpen
  \bibfield  {author} {\bibinfo {author} {\bibfnamefont {A.}~\bibnamefont
  {Serenelli}}, \bibinfo {author} {\bibfnamefont {S.}~\bibnamefont {Basu}},
  \bibinfo {author} {\bibfnamefont {J.~W.}\ \bibnamefont {Ferguson}}, \ and\
  \bibinfo {author} {\bibfnamefont {M.}~\bibnamefont {Asplund}},\ }\href
  {\doibase 10.1088/0004-637X/705/2/L123} {\bibfield  {journal} {\bibinfo
  {journal} {Astrophys. J.}\ }\textbf {\bibinfo {volume} {705}},\ \bibinfo
  {pages} {L123} (\bibinfo {year} {2009})},\ \Eprint
  {http://arxiv.org/abs/0909.2668} {arXiv:0909.2668 [astro-ph.SR]} \BibitemShut
  {NoStop}%
\bibitem [{\citenamefont {Roberts}\ \emph {et~al.}(2016)\citenamefont
  {Roberts}, \citenamefont {Dzuba}, \citenamefont {Flambaum}, \citenamefont
  {Pospelov},\ and\ \citenamefont {Stadnik}}]{Roberts:2016xfw}%
  \BibitemOpen
  \bibfield  {author} {\bibinfo {author} {\bibfnamefont {B.~M.}\ \bibnamefont
  {Roberts}}, \bibinfo {author} {\bibfnamefont {V.~A.}\ \bibnamefont {Dzuba}},
  \bibinfo {author} {\bibfnamefont {V.~V.}\ \bibnamefont {Flambaum}}, \bibinfo
  {author} {\bibfnamefont {M.}~\bibnamefont {Pospelov}}, \ and\ \bibinfo
  {author} {\bibfnamefont {Y.~V.}\ \bibnamefont {Stadnik}},\ }\href {\doibase
  10.1103/PhysRevD.93.115037} {\bibfield  {journal} {\bibinfo  {journal} {Phys.
  Rev.}\ }\textbf {\bibinfo {volume} {D93}},\ \bibinfo {pages} {115037}
  (\bibinfo {year} {2016})},\ \Eprint {http://arxiv.org/abs/1604.04559}
  {arXiv:1604.04559 [hep-ph]} \BibitemShut {NoStop}%
\bibitem [{\citenamefont {Aprile}\ \emph
  {et~al.}(2014{\natexlab{c}})\citenamefont {Aprile} \emph
  {et~al.}}]{Aprile:2012vw}%
  \BibitemOpen
  \bibfield  {author} {\bibinfo {author} {\bibfnamefont {E.}~\bibnamefont
  {Aprile}} \emph {et~al.} (\bibinfo {collaboration} {XENON100}),\ }\href
  {\doibase 10.1016/j.astropartphys.2013.10.002} {\bibfield  {journal}
  {\bibinfo  {journal} {Astropart. Phys.}\ }\textbf {\bibinfo {volume} {54}},\
  \bibinfo {pages} {11} (\bibinfo {year} {2014}{\natexlab{c}})},\ \Eprint
  {http://arxiv.org/abs/1207.3458} {arXiv:1207.3458 [astro-ph.IM]} \BibitemShut
  {NoStop}%
\bibitem [{\citenamefont {Lang}\ \emph {et~al.}(2016)\citenamefont {Lang},
  \citenamefont {McCabe}, \citenamefont {Reichard}, \citenamefont {Selvi},\
  and\ \citenamefont {Tamborra}}]{Lang:2016zhv}%
  \BibitemOpen
  \bibfield  {author} {\bibinfo {author} {\bibfnamefont {R.~F.}\ \bibnamefont
  {Lang}}, \bibinfo {author} {\bibfnamefont {C.}~\bibnamefont {McCabe}},
  \bibinfo {author} {\bibfnamefont {S.}~\bibnamefont {Reichard}}, \bibinfo
  {author} {\bibfnamefont {M.}~\bibnamefont {Selvi}}, \ and\ \bibinfo {author}
  {\bibfnamefont {I.}~\bibnamefont {Tamborra}},\ }\href {\doibase
  10.1103/PhysRevD.94.103009} {\bibfield  {journal} {\bibinfo  {journal} {Phys.
  Rev.}\ }\textbf {\bibinfo {volume} {D94}},\ \bibinfo {pages} {103009}
  (\bibinfo {year} {2016})},\ \Eprint {http://arxiv.org/abs/1606.09243}
  {arXiv:1606.09243 [astro-ph.HE]} \BibitemShut {NoStop}%
\bibitem [{Note1()}]{Note1}%
  \BibitemOpen
  \bibinfo {note} {After initial submission of the manuscript, a preprint of
  the measurement became available on arXiv~\cite
  {Akerib:2017hph}.}\BibitemShut {Stop}%
\bibitem [{\citenamefont {Szydagis}\ \emph {et~al.}(2011)\citenamefont
  {Szydagis}, \citenamefont {Barry}, \citenamefont {Kazkaz}, \citenamefont
  {Mock}, \citenamefont {Stolp}, \citenamefont {Sweany}, \citenamefont
  {Tripathi}, \citenamefont {Uvarov}, \citenamefont {Walsh},\ and\
  \citenamefont {Woods}}]{Szydagis:2011tk}%
  \BibitemOpen
  \bibfield  {author} {\bibinfo {author} {\bibfnamefont {M.}~\bibnamefont
  {Szydagis}}, \bibinfo {author} {\bibfnamefont {N.}~\bibnamefont {Barry}},
  \bibinfo {author} {\bibfnamefont {K.}~\bibnamefont {Kazkaz}}, \bibinfo
  {author} {\bibfnamefont {J.}~\bibnamefont {Mock}}, \bibinfo {author}
  {\bibfnamefont {D.}~\bibnamefont {Stolp}}, \bibinfo {author} {\bibfnamefont
  {M.}~\bibnamefont {Sweany}}, \bibinfo {author} {\bibfnamefont
  {M.}~\bibnamefont {Tripathi}}, \bibinfo {author} {\bibfnamefont
  {S.}~\bibnamefont {Uvarov}}, \bibinfo {author} {\bibfnamefont
  {N.}~\bibnamefont {Walsh}}, \ and\ \bibinfo {author} {\bibfnamefont
  {M.}~\bibnamefont {Woods}},\ }\href {\doibase 10.1088/1748-0221/6/10/P10002}
  {\bibfield  {journal} {\bibinfo  {journal} {JINST}\ }\textbf {\bibinfo
  {volume} {6}},\ \bibinfo {pages} {P10002} (\bibinfo {year} {2011})},\ \Eprint
  {http://arxiv.org/abs/1106.1613} {arXiv:1106.1613 [physics.ins-det]}
  \BibitemShut {NoStop}%
\bibitem [{\citenamefont {Lenardo}\ \emph {et~al.}(2015)\citenamefont
  {Lenardo}, \citenamefont {Kazkaz}, \citenamefont {Manalaysay}, \citenamefont
  {Mock}, \citenamefont {Szydagis},\ and\ \citenamefont
  {Tripathi}}]{Lenardo:2014cva}%
  \BibitemOpen
  \bibfield  {author} {\bibinfo {author} {\bibfnamefont {B.}~\bibnamefont
  {Lenardo}}, \bibinfo {author} {\bibfnamefont {K.}~\bibnamefont {Kazkaz}},
  \bibinfo {author} {\bibfnamefont {A.}~\bibnamefont {Manalaysay}}, \bibinfo
  {author} {\bibfnamefont {J.}~\bibnamefont {Mock}}, \bibinfo {author}
  {\bibfnamefont {M.}~\bibnamefont {Szydagis}}, \ and\ \bibinfo {author}
  {\bibfnamefont {M.}~\bibnamefont {Tripathi}},\ }\href {\doibase
  10.1109/TNS.2015.2481322} {\bibfield  {journal} {\bibinfo  {journal} {IEEE
  Trans. Nucl. Sci.}\ }\textbf {\bibinfo {volume} {62}},\ \bibinfo {pages}
  {3387} (\bibinfo {year} {2015})},\ \Eprint {http://arxiv.org/abs/1412.4417}
  {arXiv:1412.4417 [astro-ph.IM]} \BibitemShut {NoStop}%
\bibitem [{\citenamefont {Yoo}\ and\ \citenamefont
  {Jaskierny}(2015)}]{Yoo:2015yza}%
  \BibitemOpen
  \bibfield  {author} {\bibinfo {author} {\bibfnamefont {J.}~\bibnamefont
  {Yoo}}\ and\ \bibinfo {author} {\bibfnamefont {W.~F.}\ \bibnamefont
  {Jaskierny}},\ }\href {\doibase 10.1088/1748-0221/10/08/P08011} {\bibfield
  {journal} {\bibinfo  {journal} {JINST}\ }\textbf {\bibinfo {volume} {10}},\
  \bibinfo {pages} {P08011} (\bibinfo {year} {2015})},\ \Eprint
  {http://arxiv.org/abs/1508.05903} {arXiv:1508.05903 [physics.ins-det]}
  \BibitemShut {NoStop}%
\bibitem [{\citenamefont {Griest}\ and\ \citenamefont
  {Seckel}(1991)}]{Griest:1990kh}%
  \BibitemOpen
  \bibfield  {author} {\bibinfo {author} {\bibfnamefont {K.}~\bibnamefont
  {Griest}}\ and\ \bibinfo {author} {\bibfnamefont {D.}~\bibnamefont
  {Seckel}},\ }\href {\doibase 10.1103/PhysRevD.43.3191} {\bibfield  {journal}
  {\bibinfo  {journal} {Phys. Rev.}\ }\textbf {\bibinfo {volume} {D43}},\
  \bibinfo {pages} {3191} (\bibinfo {year} {1991})}\BibitemShut {NoStop}%
\end{thebibliography}%

\newpage 
\onecolumngrid

\begin{center}
{\bf SUPPLEMENTAL MATERIAL}
\end{center}

\section{Monte Carlo Simulation for DM-reflection inside the sun}

In this section, we provide details on the Monte Carlo program that
simulates the reflection of the DM particles from solar electrons. Our
program derives both the spectrum of reflected DM  and the
magnitude of the reflected flux. The simulated flux is then used to
derive the expected signal in dark matter detectors along the
procedure described in the main text with additional details provided in the subsequent sections.

Our starting point is the standard Maxwell-Boltzmann velocity
distribution for galactic DM as it standardly used in direct detection
analyses. Irrespective of the details of the assumed galactic velocity
distribution, once DM reaches the sun, its velocity will become
dominated by the sun's gravity.
One of the variables in the simulation is the impact parameter $\rho$
of an incoming DM particle. In the absence of gravitational focussing
the relevant range of $\rho$ is limited from above by the solar radius
$R_{\odot}$. The simulation scans the range of impact parameters
$0\leq \rho \leq 4R_\odot$; larger impact parameters do not change the
result as the majority of DM particles in that case misses the Sun.
We have numerically checked that the resulting magnitude of the
reflected flux and the energy spectrum of the DM coming leaving the
Sun do not change if we change the range of the impact parameter from
4$R_{\odot}$ to 3$R_{\odot}$. The initial conditions of each DM
particle are generated by randomly sampling the velocity from the
 Maxwell-Boltzmann distribution and by choosing the impact
parameter evenly from a disc with the radius of 4$R_{\odot}$.

When a DM particle is outside the Sun, we calculate its trajectory
analytically following Newton's law.
Using the classical trajectory we determine the incident angle and
velocity at the surface of the Sun. Once the DM particle is inside the
Sun, for a given cross section, we calculate the mean free path
$l_{\rm fp}(r)$ of the DM particle at each radial location $r$ inside the Sun,
\beq 
l_{\rm fp}(r) = [ n_e(r) \langle \sigma_e  v_r \rangle ]^{-1}\times v_{\rm DM} \ ,  
\eeq 
where $v_r$ is the relative velocity of the DM particle and the electron and $v_{\rm DM}$ is the velocity of the DM matter particle. 
The input used for this
calculation (the electron density inside the Sun $n_e$) is obtained from the
standard solar model of Bahcall~\cite{Bahcall:2004pz},
with competing solar models such as~\cite{Serenelli:2009yc} yielding
identical results (see Fig.~\ref{fig1}).
We then compare
$0.1\times l_{\rm fp}$ with $l_0 \equiv 0.01\times R_{\odot}$, and
choose the smaller  to be the step size ($l_{\rm step}$) for
tracing the trajectory inside the Sun. The probability of DM particle
to scatter with an electron within one step is \beq P_c = 1 - e^{-
  l_{\rm step}/l_{\rm fp}} \ .  \eeq Then we generate a number $\xi$
with a flat random distribution from 0 to 1. If $\xi< P_c$, there is no
scattering and we move the DM particle to the end point of this step
while changing its velocity according to the gravitational potential.

If $\xi> P_c$ the DM particle scatters with an electron within this
step. We randomly generate the initial energy and momentum of the
electron according to the local temperature of the Sun. Then we boost
the DM particle and the electron to their center-of-mass frame. Then
following the differential cross section we randomly generate the
directions of the outgoing DM particle and the electron. Then we boost
them back to the solar frame. (The interactions considered in this
paper are contact-type, as the mass of the mediator particle is
assumed to be larger than maximum momentum transfer. This simplifies
the distribution of the final state momenta in the collision.)

At each step we monitor if the DM particle leaves the Sun. Once it is
out of the Sun, we calculate its kinetic energy plus the potential
energy from the solar gravity and then put it into a histogram. Since
the initial impact parameter is from 0 to $4 R_{\odot}$, there is a
chance that the DM particle never passes through the Sun. If this
happens we put the initial kinetic energy of this particle into the
histogram. Then we normalize the histogram to get a normalized
distribution of the energy spectrum of solar reflected DM, which is
$F_{A_\rho}(E)$ above Eq.~(3) in the paper.

The normalized histograms for $m_{\rm DM} = 3$ MeV for different cross
sections are shown in Fig. 3 in the paper. For larger values of
$\sigma_e$, the DM particle prefers to collide with the electrons in
the outer layers of the Sun, and therefore the energy it acquires from
the Sun is relatively smaller due to lower temperatures. Whereas in
the case of smaller cross sections, the DM particle can penetrate
deeper and acquire larger energy through collisions with hotter
electrons. This explains why the red curve in the Fig. 3 drops earlier
compared to the rest.

A brief summary of the main features of the reflected flux of the DM
particles is shown in Tab.~I. For each value of the dark matter mass
and scattering cross section the average energy of the reflected DM,
the endpoint of the reflected DM spectrum (defined as the upper limit
of the energy interval containing 95\% of the reflected flux), and the
total value of the reflected DM flux at the earth position are
shown. The endpoint energy in the reflected spectrum should be
compared with the galactic endpoint, $m_{\rm DM} v_{\rm esc}^2 / 2$,
where $v_{\rm esc}$ is the escape velocity, quantifying the hardening
of the reflected spectrum. This ratio is found to be in the range
100--7000 for the parameters listed in the table.

\begin{table}[tb]
\begin{center}
\begin{tabular}{cccccc}
\hline
~&~ $10^{-38}~{\rm cm}^2$  ~&~ $10^{-37}~{\rm cm}^2$ ~&~ $10^{-36}~{\rm cm}^2$ ~&~ $10^{-35}~{\rm cm}^2$ ~&~ $10^{-34}~{\rm cm}^2$ \\
\hline
\hline
0.1 MeV~&~297 eV ~&~284 eV~&~240 eV~&~204 eV~&~153 eV \\
             ~&~1010 eV~&~967 eV~&~822 eV~&~749 eV~&~596 eV \\
             ~&~ $61~{\rm cm^{-2} sec^{-1}}$~&~ $425~{\rm cm^{-2} sec^{-1}}$~&~ $1287~{\rm cm^{-2} sec^{-1}}$~&~ $2372~{\rm cm^{-2} sec^{-1}}$~&~ $3236~{\rm cm^{-2} sec^{-1}}$ \\
 \hline
0.2 MeV ~&~433 eV~&~414 eV~&~347 eV~&~276 eV~&~197 eV \\
              ~&~1483 eV~&~1406 eV~&~1179 eV~&~979 eV~&~717 eV \\
              ~&~ $30.4~{\rm cm^{-2} sec^{-1}}$~&~ $213~{\rm cm^{-2} sec^{-1}}$~&~ $646~{\rm cm^{-2} sec^{-1}}$~&~ $1191~{\rm cm^{-2} sec^{-1}}$~&~ $1628~{\rm cm^{-2} sec^{-1}}$ \\
 \hline
0.5 MeV ~&~527 eV~&~510 eV~&~437 eV~&~339 eV~&~236 eV \\
              ~&~1796 eV~&~1725 eV~&~1466 eV~&~1165 eV~&~819 eV \\
              ~&~ $12~{\rm cm^{-2} sec^{-1}}$~&~ $84~{\rm cm^{-2} sec^{-1}}$~&~ $256~{\rm cm^{-2} sec^{-1}}$~&~ $472~{\rm cm^{-2} sec^{-1}}$~&~ $645~{\rm cm^{-2} sec^{-1}}$ \\
 \hline
2 MeV ~&~364 eV~&~373 eV~&~370 eV~&~319 eV~&~243 eV \\
            ~&~1223 eV~&~1245 eV~&~1271 eV~&~16071152 eV~&~871 eV \\
            ~&~ $2.9~{\rm cm^{-2} sec^{-1}}$~&~ $21~{\rm cm^{-2} sec^{-1}}$~&~ $63~{\rm cm^{-2} sec^{-1}}$~&~ $116~{\rm cm^{-2} sec^{-1}}$~&~ $157~{\rm cm^{-2} sec^{-1}}$ \\
 \hline
4 MeV ~&~248 eV~&~270 eV~&~314 eV~&~306 eV~&~255 eV \\
            ~&~810 eV~&~883 eV~&~1108 eV~&~1158 eV~&~934 eV \\
            ~&~ $1.3~{\rm cm^{-2} sec^{-1}}$~&~ $9.6~{\rm cm^{-2} sec^{-1}}$~&~ $30~{\rm cm^{-2} sec^{-1}}$~&~ $55~{\rm cm^{-2} sec^{-1}}$~&~ $74~{\rm cm^{-2} sec^{-1}}$ \\
 \hline
\end{tabular}
\label{tableI}
\end{center}
\caption{Features of the reflected DM flux on the earth. For each
  value of the DM mass the first and second row show the average
  energy and the end point energy of the reflected DM flux; the third
  row is the flux of DM at the surface of the earth. }
\end{table}

\begin{figure}
  \includegraphics[width=0.45\textwidth]{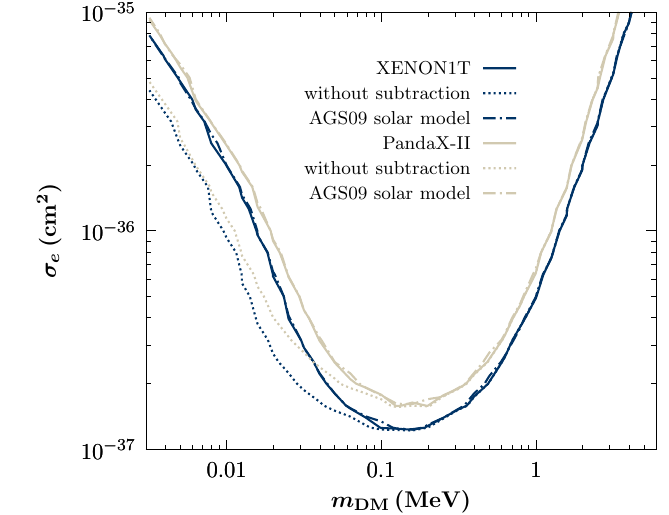}
  \caption{\footnotesize Impact of various systematic theoretical
    uncertainties on the direct detection limits originating from the
    solar reflection process as well as from the evaluation of the
    atomic scattering cross section.}
\label{fig1}
\end{figure}

\section{Evaluation of the DM electron scattering cross section}

Here we provide the details of our evaluation of the
DM($\chi$)-electron$(e)$ scattering cross section. We treat the electron recoil nonrelativistically, 
$E_e' = m_e + E_{R,e}$ with
$ E_{R,e} \ll m_e$, but allow for the general case when the incoming DM
particle may be relativistic. From the definition of momentum transfer
$\vec q = \vec p_{\chi} - \vec p_{\chi}'$,  energy conservation gives the total amount of energy lost by 
DM in the collision and hence deposited in the detector,
\begin{align}
  E_{\rm dep.} = E_{\chi} - E_{\chi}' = m_{\chi} \left[ \sqrt{1 + \frac{\vec p_{\chi}^2}{m_{\chi}^2}} -  \sqrt{1 + \frac{ |\vec p_{\chi} -\vec q |^2}{m_{\chi}^2}} \right] . 
\end{align}
The scattering cross section on bound electrons is conveniently
normalized to the non-relativistic cross section $\sigma_e$ on a free
electron (see, \textit{e.g.}~\cite{Essig:2011nj}),
$ \sigma_e \equiv {\mu_{\rm DM,e}^2 \overline{|M(q=\al m_e,E_{\chi}\to m_{\chi})|}^2}/{(16\pi
  m_{\rm DM}^2 m_e^2)},$
where the square of scattering amplitude  is summed (averaged) over intial
(final) state spins. The latter is evaluated at a momentum transfer
characteristic for an atomic process, $ q = |\vec q| = \alpha m_e$,
with any remaining $q$-dependence and/or $\chi$-energy dependence
absorbed by a DM form factor $F_{\rm DM}$,
$\overline {|M(q,E_{\chi})|}^2 =\overline{ |M( q=\al m_e, E_{\chi}\to
  m_{\chi})|}^2 \times |F_{\rm DM}( q, E_{\chi})|^2$. For the purpose
of this paper, we consider the simple case of a contact interaction
for which $F_{DM} = 1$ while more general cases are obtained in a
straightforward manner.

The differential electron recoil rate resulting from ionization of
atomic state with principal and angular quantum numbers $n$ and $l$ can be brought into the form,
\begin{align}
  \label{eq:dsigdlnErRELAT}
  \frac{ d \sigma_{nl} v }{ d\ln E_{R,e}  }  &  =   \frac{\sigma_e }{16\pi  }
   \frac{m_{\chi}^2}{\mu_{\chi e}^2 }  \int d\Omega_{\vec p_e'} \frac{d^3\vec q }{E_{\chi}  E_{\chi}'}   \, 
  |F_{\rm DM}(q,E_{\chi})|^2  |f_{nl}(q)|^2 \delta(E_{\rm dep.} - \Delta E_e)
\end{align}
where $\Delta E_e = E_{R,e} + |E_B|$. 
The $\delta$-function can be
used to perform the angular part of the integral in $d^3\vec q$%
,
\begin{align}
\frac{  d \sigma_{nl} v }{ d\ln E_{R,e} } =  
 \frac{ \bar \sigma_e }{8 \mu_{\chi e}^2}  \frac{m_\chi^2}{p_{\chi} E_{\chi}}
  \int dq d\Omega_{\vec p_e'}\, q |f_{nl}(q)|^2  |F_{\rm DM}(q,E_{\chi}) |^2 .
\end{align}
A factor $1/v$ in the usual
expression of galactic DM-electron scattering is being replaced by the
more general factor $m_{\chi}^2/(p_{\chi}E_{\chi})$.
The ionization form factor is given by, 
\begin{align}\label{eq:fion}
  |f_{nl}({q, p_e'})|^2 = \frac{p_e'}{\pi^2 q} \int_{|p_e'-q|}^{p_e'+q} dp' \, p' \sum_{m=-l}^l
  | \langle \vec p_e' | e^{ i \vec q\cdot \vec r} | nlm
  \rangle |^2 ,
\end{align}
where $\vec p' \equiv \vec p_e' - \vec q $.  For very light DM,
$|\vec q \cdot \vec r |\ll 1$ may be attained in the evaluation
of~(\ref{eq:fion}). If the final state wave function 
is approximated by a plane or Coulomb wave, and not the exact atomic electron wave function, 
the bound and scattering states are not guaranteed to be orthogonal, $ \langle nlm | \vec p_e' \rangle \neq 0$. 
To avoid spurious contributions in the
approximations employed we modify the transition matrix element by subtracting 
the unity operator,
\begin{align}
  \langle \vec p_e' | e^{ i \vec q\cdot \vec r}  | nlm \rangle \to   \langle \vec p_e' | e^{ i \vec q\cdot \vec r} - 1 | nlm \rangle
= \chi_{nl}(p') Y_{lm}(\hat p') -  \chi_{nl}(p_e') Y_{lm}(\hat p_e') 
\end{align}
where $\chi_{nl}$ is the
fourier transform of the radial Hartree-Fock wave function $R_{nl}$,
$ \chi_{nl}(q) = 4\pi (-1)^l i^l \int dr\, r^2 R_{nl}(r) j_l(q r) .$
Explicitly, one finds,
\begin{align}
\label{eq:subformfac}
\int  d\Omega_{\vec p_e'}\   
 \sum_{m=-l}^l  |\langle \vec p_e' | e^{ i \vec q\cdot \vec r} - 1 | nlm \rangle   |^2 & =   \frac{2l+1}{2 p_e' q} \int^{p_e'+q}_{|p_e' -q|} dp'\ p' |\chi_{nl}(p')|^2  %
 \nonumber \\ & + (2l+1) |\chi_{nl}(p_e')|^2 -  (-1)^l \frac{2l+1}{ 2\pi }  \chi_{nl}(p_e') \int d\Omega_{\vec p_e'}  \ \chi_{nl}(p')   P_l(\hat p' \cdot \hat p_e') . 
\end{align}
The first term on the right hand side is the DM-electron scattering
form factor previously obtained in the literature~\cite{Essig:2011nj} while
the subsequent terms originate from the subtraction.  The remaining angular
integral is evaluated as,
\begin{align}
   \int d\Omega_{\vec p_e'}  \ \chi_{nl}(p')  P_l(\hat p' \cdot \hat p_e') = 2\pi \int_{-1}^{+1} d\cos\theta\  \chi_{nl} \left( \sqrt{p_e'^2 + q^2 - 2 p_e' q \cos\theta}  \right)  
  P_l \left( \frac{p_e' -  q \cos\theta }{ \sqrt{p_e'^2 + q^2 - 2 p_e' q \cos\theta}}  \right) .
\end{align}
Since two unit vectors are dotted as argument of the Legendre polynomial $P_l$,
the latter integrand is always well-behaved. One can verify that
the subtraction works by noting that,
\begin{align}
   \frac{2l+1}{2 p_e' q} \int^{p_e'+q}_{|p_e' -q|} dp'\ p' |\chi_{nl}(p')|^2 \to  (2l+1) |\chi_{nl}(p_e')|^2\qquad (q\to 0),
\end{align}
and that $P_l(1)=1$ so that Eq.~(\ref{eq:subformfac}) indeed vanishes
identically for $q\to 0$. We have also verified this numerically in
our computer code.  It turns out that the subtraction has only
a relatively minor influence on the derived bounds, see Fig.~\ref{fig1}. We
note in passing that the case considered here is quite different from the case when DM is
slow and $|\vec q \cdot \vec r| \gg 1$ for which the ionization process
becomes short-distance dominated, and a relativistic atomic treatment
becomes necessary~\cite{Roberts:2016xfw}.

The minimum incoming momentum to produce an electron recoil $E_{R,e}$
is obtained from the $\delta$-function in (\ref{eq:dsigdlnErRELAT})
when $\vec q$ and $\vec p_{\chi}$ are parallel,
\textit{i.e.}~$\cos\theta_{ q p_{\chi}} = 1$,
\begin{align}
\label{eq:pmin}
  p_{\chi}^{\rm min} = \frac{q}{2 (1-\Delta E_e^2/q^2)} \left[ 1 - \frac{\Delta E_e^2}{q^2}
 + \frac{\Delta E_e}{q} \sqrt{ \left( 1 - \frac{\Delta E_e^2}{q^2} \right) 
 \left( 1 +  \frac{4 m_{\chi}^2}{q^2} - \frac{\Delta E_e^2}{q^2}   \right)     }   
  \right].
\end{align}
This expression is exact and is used in the integral that computes the
average over the incoming energy spectrum (see main text). In the
non-relativistic limit, expanding in $\Delta E_e/q \ll 1$ one recovers
the expression for $v_{\rm min} = p_{\chi}^{\rm min} / m_{\chi}$ given
in previous works~\cite{Essig:2011nj}.

\section{Modeling of LXE detector response}

Here we describe a simple procedure that is aimed at capturing the
dominating factors in the detection of scintillation (S1) and
ionization (S2) signals following an electron recoil in a LXE
detector. Our treatment largely follows previous experimental,
theoretical, and joint theory-experiment
studies~\cite{Aprile:2012vw,Lang:2016zhv,McCabe:2017rln}.  With the
results of the paper together with the details of the MC simulation
provided in the previous section of the supplement, we invite the
experimental collaborations to perform their own dedicated analysis of
the reflected dark matter signal.

Given a total energy deposition $E_{\rm dep.} = E_{\chi} - E_{\chi}'$
from DM scattering on an atomic electron, the average number of
produced quanta at the interaction point~is,
\begin{align}
  \langle N_Q \rangle = \frac{E_{\rm dep.}}{W}
  =   E_{\rm dep.}  L_y + E_{\rm dep.} Q_y
  = \langle n_{\gamma} \rangle + \langle n_e \rangle  ,
\end{align}
with $W = 13.7\, \eV$~\cite{Dahl:2009nta,Akerib:2016qlr}.
The quanta are partitioned into $n_e$ ionized electrons escaping the
interaction point and $n_\gamma$ scintillation photons; $Q_y$ and
$L_y$ denote the energy-dependent charge and light yields,
respectively.  For the purpose of setting limits we only use data
above $E_{\rm dep.} = 0.19\,\keV$ in the computation of
$ \langle n_e \rangle$, corresponding to the lowest energy at which
$Q_y$ was measured~\cite{Xe127}.%
\footnote{After  initial submission of the manuscript, a preprint of the
  measurement became available on arXiv~\cite{Akerib:2017hph}.} In addition,
data on $Q_y$ from \cite{Akerib:2015wdi} is used; see
also~\cite{Szydagis:2013sih,Goetzke:2016lfg}.  The light output is
then obtained self-consistently by demanding conservation of~$N_Q$,
\begin{align}
  L_y = \frac{1}{W} - Q_y .
\end{align}
For $E_{\rm dep.} \lesssim 10$~keV the charge and scintillation yields
depend only mildly on the applied drift
voltage~\cite{Szydagis:2011tk,Szydagis:2013sih,Lenardo:2014cva}; for
given $E_{\rm dep.}$, $Q_y$ ($L_y$) varies by about $ 10\%$ in the
range $120-730$~V/cm, \textit{i.e.}~in the range of applied
drift-fields accross the various experiments.  We hence neglect this
experiment-specific detail as it is expected to have only relatively
minor impact on the resulting bounds.
Finally, heat losses lead to a quenching in the nuclear recoil signal,
and subsequently to a fluctuation in~$ N_Q $. For electron recoils,
heat losses are negligible~\cite{Szydagis:2011tk,Szydagis:2013sih} and
we take the number of produced quanta for a given deposited energy as
a constant, $ N_Q = \langle N_Q \rangle$.

\subsection{Fluctuations at the interaction point}

Recombination at the interaction point shifts the partition of
$n_{\gamma}$ and $n_e$ while holding their sum
$N_Q = n_{\gamma} + n_e$ fixed. The
quantities $n_{\gamma}$ and $ n_e$ are the end products after an
initial number of ions $n_i$ (yielding electrons) and excitons
$n_{ex}$ (yielding scintillation) had been created but were
redistributed because of recombination described by parameter $ r $,
$  n_e = n_i (1-r)$, $n_{\gamma}  =  n_i (r + \alpha) ,$
with $\alpha \equiv n_{ex}/n_i$. Note that
$ N_Q = n_{\gamma} + n_e = n_{i} + n_{ex} $. Fluctuations in $r$
itself, leading to an observed variance that is in excess from one
that is expected from a binomial process, are most important for
larger energy depositions
$E_R \gtrsim 2\,\keV$~\cite{Akerib:2015wdi,Akerib:2016qlr}; see also
\cite{Lenardo:2014cva}. Since the bulk of the solar flux is
energetically lower, we neglect this complication. Thus, we
follow~\cite{Lang:2016zhv} and approximate the primary signal
formation by directly computing the probabilities of producing either
$n_{\gamma}$ or~$n_{e}$ as,
\begin{align}
  P(n_{e,\gamma} | \langle n_{e,\gamma} \rangle ) = \binomial (n_{e,\gamma} |  N_Q , f_{e,\gamma} ) %
\end{align}
where, $ f_{e,\gamma} = n_{e,\gamma}/N_Q$, such that
$f_e + f_{\gamma} = 1 $.
Since $\sigma_{e,\gamma}^2 = f_{e,\gamma} (1-f_{e,\gamma}) N_Q $ it follows that
$\sigma_{\gamma}^2 = \sigma_{e}^2 $ and  $  P(n_e | \langle n_e \rangle ) =    P(n_{\gamma} | \langle n_{\gamma} \rangle )$. 

\subsection{Detector-specific fluctuations}
\label{sec:detect-spec-fluct}

The S1 signal is obtained from the probability of detecting $n_{\PE}$
photons from $n_{\gamma}$ produced, and is modelled by a binomial
distribution with overall light collection efficiency~$g_1$ (the
experiment-specific values of $g_1$ are found in the main text),
\begin{align}
  P(n_{\PE} | n_{\gamma}) = \binomial (n_{\PE} |n_{\gamma}, g_1 ).
\end{align}

For computing the S2 signal, one needs to account for the survival
probability $p_{\rm surv}$ of electrons when they are drifted by a
distance $\Delta z$ until the liquid-gas interface with a ballpark
velocity $v_d \sim 1.7 \rm mm/\mu\sec$~\cite{Aprile:2012vw,
  Yoo:2015yza},
\begin{align}
  \label{eq:psurv}
  p_{\rm surv} = \exp \left( - \frac{\Delta z}{\tau v_{d}} \right). 
\end{align}
The electron lifetime varies across experiments.  We follow the
reasoning presented in~\cite{Lang:2016zhv} and assume that the
electron lifetime is distributed uniformly over $[0, 2/3]$~mm/$\mu$s.
The probability of $n_{e}^{\rm surv}$ electrons reaching the gas-phase
of the detector is then found from the compound distribution function,
where one marginalizes over the production location,
\begin{align}
  P(n_e^{\rm surv} | n_e) = 
  \frac{3}{2} \int_0^{2/3} 
 d \left( \frac{\Delta z}{\tau} \right)\, \binomial(n_e|  p_{\rm surv}(\Delta z/\tau) ) .
\end{align}
To relax the numerical demand, in the actual analysis we assign an
average electron survival probability
$ P(n_e^{\rm surv} | n_e) \simeq \langle p_{\rm surv}\rangle = 0.8 $
obtained from the expectation value of~(\ref{eq:psurv}) which then
allows to perform the sum in (\ref{eq:pdfs2}) below.

In a final step we account for the PMT resolution, using as a
representative value
$\sigma_{\rm PMT} \simeq 0.4 \sqrt{n_{\rm PE}}$ for the
low-photon count in S1. For S2, once the electron reaches the gas
phase it receives a gain factor $g_2$ in the number of produced
photo-electrons that are detected (for the experiment-specific values
of $g_2$ see main text). The process is modeled by a Gaussian with
representative standard deviation
$\sigma_{S2} = 7 \sqrt{n_e^{\rm surv}}$~\cite{Aprile:2013blg}. The
respective probabilities of detection read,
\begin{align}
  P(S1|n_{\PE}) & = \gaussian(S1| n_{\PE}, \sigma_{\rm PMT}), \\
  P(S2|n_e^{\rm surv}) & = \gaussian(S2| g_2 n_{e^{\rm surv}}, \sigma_{S2}) . 
\end{align} 

Collecting all factors allows one to estimate the correlated PDF for
observing S1 and S2, given an energy deposition of $E_{\rm dep.}$,
\begin{align}
\label{eq:pdfs1s2}
  P(S1,S2 | E_R) = \sum_{n_e^{\rm surv}}\sum_{n_{\PE}}  \sum_{n_{\gamma}} 
  P(S2|n_e^{\rm surv})  P(S1|n_{\PE})  P(n_e^{\rm surv} | n_e)   P(n_{\PE} | n_{\gamma})
  P(n_{\gamma} | \langle n_{\gamma} \rangle ) . 
\end{align}
Such expression for the joint PDF in S1 and S2 makes it amenable to a
likelihood analysis of the experimental data.
For the purpose of this paper, where we primarily explore the
principal experimental sensitivity and do not intend to forestall a
dedicated experimental analysis, we consider the signals S1 and S2
separately, leaving an analysis in the correlated signal for more
in-depth work.
The pdf for S1 is then obtained from,
\begin{align}
\label{eq:pdfs1}
  P(S1| E_R) = \sum_{n_{\PE}}  \sum_{n_{\gamma}}  P(S1|n_{\PE}) P(n_{\PE} | n_{\gamma})
  P(n_{\gamma} | \langle n_{\gamma} \rangle ) = \sum_{n_{\PE}}   P(S1|n_{\PE}) 
  \binomial(n_{\PE} | N_Q , f_{\PE} f_{\gamma}) ,
\end{align}
where in the last equality we have performed the sum over $n_{\gamma}$.
For obtaining the PDF in S2 only, one evaluates,
\begin{align}
\label{eq:pdfs2}
  P(S2| E_R) = \sum_{n_e^{\rm surv}}  \sum_{n_e} 
  P(S2|n_e^{\rm surv})   P(n_e^{\rm surv} | n_e) 
  P(n_e | \langle n_e \rangle ) . 
\end{align}
Of course, the PDFs (\ref{eq:pdfs1}) and (\ref{eq:pdfs2}) also follow
directly from (\ref{eq:pdfs1s2}) by marginalizing over S2 and S1,
respectively.

\section{Details on the exemplary DM model}
\label{sec:exemplary-dm-model}

The nonrelativistic scattering cross section on free electrons
obtained from the Lagrangian given in the main text is found to be
\begin{align}
\sigma_e = \frac{1}{\pi} G_{\chi e}^2 \mu^2 ,
\end{align}
where $\mu$ is the reduced mass.  In turn, the total annihliation
cross section to electrons in terms of the squared center-of-mass
energy $s$ reads,
\begin{align}
  \sigma_{\rm ann}(s)=
  \frac{G_{\chi e}^2}{12 \pi s} (s+2 m_e^2 )\sqrt{s-4m_{\chi}^2} \sqrt{s-4 m_e^2}
\end{align}
In the non-relativistic expansion, $s= 4m_{\chi}^2 + m_{\chi}^2 v$,
and one finds the velocity scaling corresponding to $p$-wave
annihilation,
\begin{align}
  \sigma_{\rm ann} v  = v^2 \times  \frac{G_{\chi e}^2}{12 \pi }  (m_e^2 + 2m_{\chi}^2) \sqrt{ 1- \frac{m_e^2}{m_{\chi}^2}} . 
\end{align}
We have calculated the relic abundance with three different methods:
semi-analytically following~\cite{Griest:1990kh}, numerically
following~\cite{Gondolo:1990dk}, and through an implementation of the
publicly available computer code Micromegas~\cite{Belanger:2006is}
with mutually agreeing results. Imposing the relic density condition,
allows one to obtain the expectation for $\sigma_e$ as a function of
$m_{\chi}$ as provided in the main text. Within this model, the area below the line 
is disfavored by  DM overproduction, while the area above the line makes 
$\chi$ to form only a fraction of the total DM energy density.

\end{document}